\documentclass[a4paper,11pt]{article}

\usepackage{jheppub} 
\usepackage[utf8]{inputenc}
\usepackage[T1]{fontenc} 
\usepackage{multirow}
\usepackage{tikz}

\usepackage{pdflscape}
\usepackage{booktabs}
\usepackage{verbatim}

\usetikzlibrary{matrix, positioning}

\usepackage{slashed}

\usepackage{subcaption}
\usepackage{dsfont}

\def\la{\langle}
\def\ra{\rangle}

\def\ep{\epsilon}

\def\ubar{\overline{u}}

\def\measure#1{\left[{\rm d}p_{#1}\right]}
\def\flm{F_\text{LM}}
\def\flv{F_\text{LV}}
\def\flvv{F_\text{LVV}}
\def\la{\big\langle}
\def\ra{\big\rangle}

\newcommand{\be}{\begin{equation}}
\newcommand{\ee}{\end{equation}}

\title{Non-factorisable contribution to $t$-channel single-top production}

\author[a]{Christian Br\o{}nnum-Hansen,}
\author[a]{Kirill Melnikov,}
\author[a]{J\'{e}r\'{e}mie Quarroz,}
\author[a,b]{Chiara Signorile-Signorile,}
\author[a]{Chen-Yu Wang}

\def\KITA{Institute for Theoretical Particle Physics, KIT, Karlsruhe, Germany}
\def\KITB{Institute for Astroparticle Physics, KIT, Karlsruhe, Germany}

\emailAdd{christian.broennum-hansen@kit.edu}
\emailAdd{kirill.melnikov@kit.edu}
\emailAdd{jeremie.quarroz@kit.edu}
\emailAdd{chiara.signorile-signorile@kit.edu}

\affiliation[a]{\KITA}
\affiliation[b]{\KITB}

\preprint{TTP22-021, P3H-22-035}

\abstract{We compute the non-factorisable ${\cal O}(\alpha_s^2)$ corrections to $t$-channel single-top 
quark production at the LHC.
These peculiar corrections arise because of interactions between  the heavy- and the light-quark lines and appear for the
very first time at next-to-next-to-leading order in perturbative QCD.
We find that the non-factorisable corrections
change the single-top production cross section and the relevant  kinematic
distributions in this process by about half a percent. 
}

\begin{document}
\maketitle
\flushbottom

\section{Introduction}

Studies of top quarks are important for the exploration of the Standard Model (SM) and in searches for its
extensions.  With a  mass of more than 170 GeV, the top quark is the heaviest
elementary particle of the SM and has an exceptionally strong coupling to the Higgs boson.  These special features
of top quarks make it plausible that they play a particular role in the underlying mechanism of electroweak
symmetry breaking and may have significant couplings to heavy New Physics.

The large top-quark mass is the reason
behind its short lifetime which, in fact, is so short that, once produced, top quarks decay before hadronising
into mesons and baryons.  This has many interesting consequences including the fact that the information about
top-quark polarisation is passed to its decay products offering an opportunity to study this aspect of QCD
without non-perturbative contamination. 
 
At the LHC, top quarks are mainly produced in pairs via strong interactions. 
Theoretical predictions for $t \bar t$  pair production are very advanced and include
next-to-leading-order (NLO) QCD~\cite{Nason:1987xz} and electroweak corrections~\cite{Kuhn:2005it}, soft gluon resummation~\cite{Bonciani:1998vc,Catani:1996dj,Beneke:2009rj,Czakon:2009zw,Cacciari:2011hy},
and total and fully differential next-to-next-to-leading-order (NNLO) QCD corrections~\cite{Czakon:2013goa,Czakon:2017wor,Behring:2019iiv,Czakon:2020qbd} in the narrow-width approximation.

Top quarks can also be produced via electroweak interactions; this mechanism is referred to as \textit{single-top production}. 
Interestingly,  rates for single-top production at the LHC are quite significant. 
In fact, the single-top quark production cross section is smaller than the $t \bar t$  production
cross section by only about a factor of four.  However, since the $pp \to t \bar t$ cross section at the LHC is large, of the order of
a nanobarn, rates for single-top production turn out to be very high as well. 
As a result, there is a well-developed experimental program for studying single-top production
at the LHC that focuses on inferring information
about the top-quark width~\cite{CMS:2014mxl}, mass~\cite{CMS:2017mpr} and
polarisation~\cite{CMS:2015cyp}, as well as using this process
to constrain  possible anomalous couplings in the $tWb$ vertex~\cite{Kane:1991bg, Gonzalez-Sprinberg:2015dea}.
Studies of single-top production are also used to constrain  the CKM matrix
element $V_{tb}$, which has been measured both at the Tevatron~\cite{CDF:2015gsg} and at 
the LHC~\cite{CMS:2014mgj}.
Finally, single-top production  can be used to provide interesting probes
of parton distribution functions (PDFs). For example,
comparison of single-top and single-anti-top production cross sections can be used to constrain  
 ratios of up- and down-quark  distribution functions 
at fairly  large values of Bjorken $x$~\cite{Alekhin:2015cza,CMS:2019jjp,ATLAS:2017rso}.

Single top quarks are produced in hadron collisions in three distinct ways that are conventionally  referred  to as channels.
 The $t$-channel production refers to a process where a $W$ boson is exchanged between two quark lines
and a top quark is produced on one of them as the result of the flavour-changing $tWb$ interaction. 
The $s$-channel production refers to a process where a virtual $W$ boson is first
created in the collisions of light quarks and later decays into a top quark and an anti-bottom quark. 
The associated production  refers to a process where an off-shell bottom quark is produced and then decays
into a top quark and a $W$ boson.

Among the three channels, $t$-channel production is responsible for about $70\%$ of the single-top
production cross section. Because of that, this contribution has been carefully scrutinised in the Standard Model.
In particular,  NLO QCD  and NNLO QCD corrections to $t$-channel single-top production were computed in 
Refs.~\cite{Bordes:1994ki,PhysRevD.70.094012,Cao:2004ky,Cao:2005pq,Harris:2002md,Schwienhorst:2010je}
and Refs.~\cite{Brucherseifer:2014ama,Berger:2016oht,Berger:2017zof,Campbell:2020fhf}, respectively.
Typically, the inclusive cross section for single-top production changes by about 2-3 percent at next-to-leading order\footnote{For certain
  parton distribution functions, the NLO QCD corrections to single-top production cross section 
  can be more significant, see e.g. Ref.~\cite{Campbell:2020fhf}.}
  and
by 1-3 percent at next-to-next-to-leading order. The reason behind the smallness of these corrections is the proximity
of single-top production and deep-inelastic scattering processes which means that a bulk of QCD corrections is absorbed into
PDFs by virtue of the fitting process. This proximity is destroyed if selection cuts are applied
to the final state that are not inclusive with respect to QCD radiation. Indeed, once this is done both NLO and NNLO
QCD corrections become larger and can reach ${\cal O}(10 \%)$ in certain kinematic distributions~\cite{Campbell:2020fhf}. 

It is interesting to remark that the above results~\cite{Brucherseifer:2014ama,Berger:2016oht,Berger:2017zof,Campbell:2020fhf}
were obtained in the so-called factorisation approximation
that neglects the dynamical crosstalk  between the two quark lines.  This was done for the following reason:
although the two incoming or outgoing quarks can interact by exchanging a gluon  already at NLO,
it is easy to see that such a contribution  does not affect the production cross section at this order because of
colour conservation. However, such non-factorisable corrections start contributing at NNLO but they
are colour-suppressed relative to factorisable contributions.
Conversely, it was recently argued~\cite{Liu:2019tuy} that these non-factorisable contributions
could be enhanced by a factor $\pi^2$ due to the
Glauber phase~\cite{glauber, Cheng:1969tje}, which would compensate for the colour suppression. In fact, explicit computations
of non-factorisable corrections performed in Ref.~\cite{Liu:2019tuy} for Higgs production in weak boson fusion proved the existence of such
an enhancement factor. 

The non-factorisable contributions are quite peculiar. Indeed, they are ultraviolet finite and thus do not require
any renormalisation. In addition,  as we will show later, they are entirely Abelian, at least at NNLO,
which  implies a remarkable simplification in their  infrared structure.
Moreover,  they do not contain  collinear singularities since, in physical gauges, collinear singularities originate from
the emission and absorption of a real or virtual gluon by the same on-shell particle and for the non-factorisable corrections
this is impossible because of their definition. 
As a consequence, all infrared divergences that may appear in non-factorisable corrections are of soft origin 
and, in dimensional
regularisation, correspond to at most double poles in the regulator.

We also expect that  virtual effects  play a more important role in non-factorisable
corrections than the real-emission contributions. 
This is because the enhancement of the non-factorisable corrections by a Glauber phase is a virtual effect that, in principle,
does not require scattering to occur and, hence, remains present also at zero momentum transfer where the cross section
is the largest. 
Indeed, no scattering means no real radiation so we expect that
real-emission contributions are, naturally, less important. 

Very recently, the two-loop non-factorisable contributions to single-top production were computed in 
Ref.~\cite{Bronnum-Hansen:2021pqc}. The results of that  reference are, however, not complete since
two additional contributions -- double-real emissions and virtual corrections to the single-real emission --
are required to compute infrared-finite cross sections and kinematic distributions.  In this paper we complete
the calculation of non-factorisable corrections to the single-top production by computing the two remaining NNLO contributions  and, for the first
time, provide physical results for non-factorisable corrections to this process. 

The paper is organised as follows.  In Sec.~\ref{sec:generalities} we
introduce the  notation and describe the set-up of the calculation.
We proceed in Sec.~\ref{sec:pole_canc} with the treatment of the
infrared singularities that affect both the real radiation and the virtual corrections.
In Sec.~\ref{sec:amp_calc} we briefly discuss  the calculation of
the real and virtual amplitudes.
Phenomenological results are reported in Sec.~\ref{sec:results}. We conclude  in Sec.~\ref{sec:conclusions}.

\begin{figure}[t]
    \centering
    \begin{subfigure}[ht]{0.3\linewidth}
        \centering
        \includegraphics[height=3.2cm]{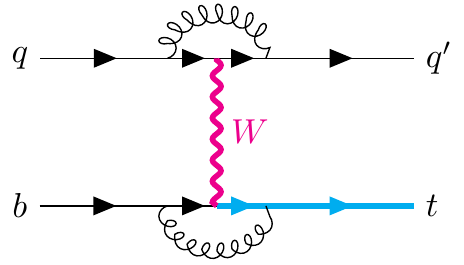}
    \end{subfigure}
    \hspace{2cm}
    \begin{subfigure}[ht]{0.3\linewidth}
        \centering
        \includegraphics[height=2.4cm]{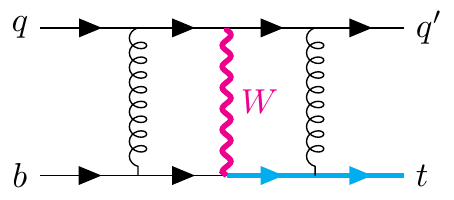}
    \end{subfigure}
    \caption{Examples of diagrams contributing to NNLO QCD corrections to  single-top production.
      The diagram on the left is part of the factorisable corrections as the two quark lines interact solely through the exchange of a colourless $W$ boson. On the right, the quark lines are additionally connected by the exchange of two gluons. We classify the latter as a non-factorisable diagram.}
    \label{fig:example}
\end{figure}

\section{Colour decomposition of non-factorisable contributions and their singular limits}
\label{sec:generalities}

We mentioned in the introduction that non-factorisable contributions
are, effectively, Abelian and that this simplifies their calculation
significantly. In this section we explain this point in detail.

\subsection{Elastic process}

We start with the discussion of
the colour decomposition of the relevant partonic processes.
Consider single-top production in the $t$-channel
\be
1_q + 2_b \to 3_{q'} + 4_t \; ,
\label{eq:process}
\ee
where by $i_f$ we refer to a parton of type $f$ with momentum $p_i$. 
Since this process is mediated by a $W$ boson, there is no colour transfer between the two fermion lines.
To make this explicit, we use  the  colour-space formalism\footnote{The colour-space
  formalism is reviewed in Ref.~\cite{Catani:1996vz}.}  and  write the Born amplitude as
\be
\label{eq:M0} 
\langle c |{\cal M}_{0}(1_q,2_b,3_{q'},4_t) \rangle  = \delta_{c_3 c_1} \delta_{c_4 c_2} \, A_0(1_q,2_b,3_{q'},4_t) \; ,
\ee
where $A_0$ is the colour-stripped amplitude.

In order to compute the NNLO QCD corrections to the process in Eq.~\eqref{eq:process}, 
we need the expression for the corresponding one- and two-loop amplitudes.
We begin  with the former  and write it as
\be
\langle c |{\cal M}_{1}(1_q,2_b,3_{q'},4_t) \rangle
= \frac{\alpha_s}{2 \pi}  \left ( \delta_{c_3 c_1} \delta_{c_4 c_2} \, 
A_{1}(1_q,2_b,3_{q'},4_t) +
t^{a}_{c_3 c_1} t^a_{c_4 c_2} \, 
B_1(1_q,2_b,3_{q'},4_t)
\right ),
\label{eq:M1}
\ee
where $A_1$ describes emissions and absorptions of virtual gluons by the same fermion
line and $B_1$  refers to  a  one-loop
amplitude that describes interactions between light- and heavy-fermion lines.
Also, $t^a_{ij}$ are matrix elements
of the $SU(3)$ generators and $\alpha_s \equiv \alpha_s(\mu)$ is the renormalised strong coupling constant in the
$\overline{\text{MS}}$ scheme (see Appendix~\ref{sec:app_A} for details).
It is this last amplitude that is of interest to us,  as it contributes to non-factorisable corrections.
The $B_1$
amplitude is ultraviolet-finite but infrared-divergent; the infrared divergence is described by the following formula
\be
\label{eq:B1_4point}
B_1(1_q,2_b,3_{q'},4_t) =    I_1(\ep) \; A_0(1_q,2_b,3_{q'},4_t) +  B_{1,\rm fin} (1_q,2_b,3_{q'},4_t)\, ,
\ee
where
\be
I_1(\ep) \equiv
I_1(1_q,2_b,3_{q'},4_t; \ep) =
\frac{1}{\ep} \left [ \log \left( \frac{ p_1 \cdot p_4 \; p_2  \cdot p_3 }{ p_1 \cdot  p_2 \; p_3 \cdot p_4 } \right) + 2\pi i \right ].
\label{eq:I1}
  \ee

\vspace*{0.3cm}
We can write the two-loop amplitude  in a similar manner.  First, we define the non-factorisable contribution to the
amplitude as follows
\be
\langle c |{\cal M}_{2}(1_q,2_b,3_{q'},4_t) \rangle  = \left ( \frac{\alpha_s}{2 \pi} \right )^2  \left ( ... +
\frac{1}{2}\{ t^a,t^b \}_{c_3 c_1} \frac{1}{2} \{ t^a,t^b \}_{c_4 c_2} \; B_2(1_q,2_b,3_{q'},4_t) 
\right ),
\ee
where ellipses stand for factorisable contributions as well as contributions that vanish upon interference with the tree-level amplitude, Eq.~\eqref{eq:M0}.
The non-factorisable amplitude $B_2$ is infrared-divergent; these divergences
can be written in the following way 
\be
\begin{split} 
  B_2(1_q,2_b,3_{q'},4_t) = 
  &-\frac{I_1^2(\ep)}{2}  \; A_0(1_q,2_b,3_{q'},4_t) + I_1(\ep)\; B_1(1_q,2_b,3_{q'},4_t)\\
  &+ B_{2,\rm fin}(1_q,2_b,3_{q'},4_t) \, .
\end{split}
\label{eq:B2}
\ee
We stress that the finite contributions to $B_2$ arise from the two last terms.
Hence, to obtain  $B_{2,\rm fin}$ in Eq.~(\ref{eq:B2}), we require the one-loop amplitude $B_1$ to $\mathcal{O}(\ep)$.

To compute the cross section, we need a particular combination  of these elastic amplitudes. We require
\be
\begin{split} 
  & |{\cal M}_1(1_q,2_b,3_{q'},4_t)|_{\rm nf}^2
  + 2 {\rm Re} \left [ {\cal M}^*_0(1_q,2_b,3_{q'},4_t) \, {\cal M}_2(1_q,2_b,3_{q'},4_t) \right ]_{\rm nf}
\\
&   = \frac{N^2-1}{4} \left (\frac{\alpha_s}{2\pi} \right )^2 \Big  [ 
  -{\rm Re} \left [ I_1^2(\ep) \right ] |A_0(1_q,2_b,3_{q'},4_t)|^2
+ | B_1(1_q,2_b,3_{q'},4_t)|^2 
  \\
  & \hspace{32mm} + 2 {\rm Re} \left [ I_1(\ep)  \, A_0^* (1_q,2_b,3_{q'},4_t) \, B_1(1_q,2_b,3_{q'},4_t) \right ]
\\
&  \hspace{32mm}   + 2 {\rm Re} \left [ A_0^*(1_q,2_b,3_{q'},4_t) \, B_{2,\rm fin}(1_q,2_b,3_{q'},4_t) \right ]
 \Big  ],
\end{split}
\label{eq:virtualpoles}
 \ee
where $N = 3$ is the number of colours. 
We can now manipulate Eq.~\eqref{eq:virtualpoles} to make all the divergences explicit and expose
terms that contribute through  ${\cal O}(\ep^0)$.  We obtain
 \be
 \label{eq:double_virt_poles}
 \begin{split} 
 & |{\cal M}_1(1_q,2_b,3_{q'},4_t)|_{\rm nf}^2
 + 2 {\rm Re} \left [ {\cal M}^*_0(1_q,2_b,3_{q'},4_t) {\cal M}_2(1_q,2_b,3_{q'},4_t) \right ]_{\rm nf}
\\
&  =
 \frac{N^2-1}{4} \left (\frac{\alpha_s}{2\pi} \right )^2 \bigg  [ 
2 \, (  {\rm Re} \left [ I_1(\ep) \right ] )^2 \,
  |A_0(1_q,2_b,3_{q'},4_t)|^2
+ | B_{1,\rm fin}(1_q,2_b,3_{q'},4_t)|^2
  \\
  & \hspace{32mm} + 4 {\rm Re} \left [ I_1(\ep)  \right]  \, {\rm Re} \left[ A_0^* (1_q,2_b,3_{q'},4_t) \,  B_{1,\rm fin} (1_q,2_b,3_{q'},4_t)  \right ]
\\
&  \hspace{32mm}
+ 2 {\rm Re} \left [ A_0^*(1_q,2_b,3_{q'},4_t) \, B_{2,\rm fin}(1_q,2_b,3_{q'},4_t) \right ]
 \bigg  ]  \; .
\end{split}
 \ee
 It follows that the first term contains a $1/\ep^2$ divergence, the third a $1/\ep$
 divergence and the remaining two terms are finite. 

\subsection{Single-real emission contributions}  

Similarly, the tree-level amplitude for the single-emission process 
\be
\label{eq:1emission_proc}
1_q + 2_b \to 3_{q'} + 4_t + 5_g \, ,
\ee
reads 
\be
\begin{split} 
  \langle c |{\cal M}_{0}(1_q,2_b,3_{q'},4_t;5_g) \rangle   = \,  g_{s,b}
  \bigg  [& t^{c_5}_{c_3 c_1}  \delta_{c_4 c_2} A^{L}_0(1_q,2_b,3_{q'},4_t; 5_g)
  \\
  & + t^{c_5}_{c_4 c_2}  \delta_{c_3 c_1} A^{H}_0(1_q,2_b,3_{q'},4_t; 5_g)
  \bigg ],
\end{split}
\label{eq:M0R}
\ee
where $A^L_{0}$ and $A^H_{0}$ are colour-stripped amplitudes 
that describe gluon emission off the light- and heavy-quark lines respectively.
Here and in the following $g_{s,b}$ is the bare coupling constant.\footnote{We stress that 
in this paper we have used $\alpha_s$ to indicate the renormalised coupling
constant and suppressed its dependence on the scale $\mu$, while $g_{s,b}$ is the bare coupling.}
The soft limits of these colour-ordered amplitudes are relevant for the construction of subtraction terms.
To describe them,   we introduce the eikonal current 
\be
J^\mu(i,j;k) =    \frac{p_i^\mu}{p_i \cdot p_k} -  \frac{p_j^\mu}{p_j \cdot p_k}\;,
\ee
and its contraction with the polarisation vector of a gluon with momentum $k$
\be
\varepsilon_{k,\mu} J^\mu(i,j;k) = J(i,j;k,\varepsilon_k)\;.
\ee
Then, we write 
\be
\begin{split} 
  S_5  \, A^{L}_0(1_q,2_b,3_{q'},4_t; 5_g)  &=  J(3,1;5,\varepsilon_5)   \, 
  A_0(1_q,2_b,3_{q'},4_t)\;, \\
  S_5  \, A^{H}_0(1_q,2_b,3_{q'},4_t; 5_g)   &=  J(4,2;5,\varepsilon_5)   \, 
  A_0(1_q,2_b,3_{q'},4_t)\;.
\end{split}
\ee
Here we have introduced the operator $S_i$, which extracts the leading
singular behaviour in the soft limit $p_i \rightarrow 0$ of the function it acts upon.

We will also need the one-loop contribution to the amplitude of the process in Eq.~\eqref{eq:1emission_proc}. 
 Its  colour  decomposition  reads 
\be
\begin{split} 
    \langle c | {\cal M}_{1}(1_q,2_b,3_{q'},4_t&;5_g) \rangle
\,  =  g_{s,b} \left ( \frac{\alpha_s}{2 \pi}  \right )  \bigg [ 
t^{c_5}_{c_3 c_1} \delta_{c_4 c_2} \, A_1^L(5_g) 
+ t^{c_5}_{c_4 c_2} \delta_{c_3 c_1} \, A_1^H(5_g)
\\
& + \frac{1}{2} \left \{ t^{a},t^{c_5} \right \}_{c_3 c_1} t^a_{c_4 c_2} \, B_1^{sL}(5_g)
+ \frac{1}{2} \left [ t^{a},t^{c_5} \right ]_{c_3 c_1} t^a_{c_4 c_2} \, B_1^{aL}(5_g)
\\
& + \frac{1}{2} \left \{ t^{a},t^{c_5} \right \}_{c_4 c_2} t^a_{c_3 c_1} \, B_1^{sH}(5_g)
+ \frac{1}{2} \left [ t^{a},t^{c_5} \right ]_{c_4 c_2} t^a_{c_3 c_1} \, B_1^{aH}(5_g)
\bigg ].
\end{split} 
\label{eq:M1R}
\ee
In Eq.~\eqref{eq:M1R}  we  split the full amplitude into colour-stripped amplitudes that describe emissions by
light- and heavy-quark lines separately.
For each of the quark lines, we have also split the amplitudes into colour-symmetric and colour-antisymmetric parts,
indicated with superscripts $s$ and $a$ respectively.
The colour-symmetric ones are purely Abelian and the colour-antisymmetric ones are sensitive to the non-Abelian nature of QCD, including
contributions due to the gluon self-coupling. Note that we have suppressed
the dependence of the amplitudes $A_1$ and $B_1$ on the quark momenta
but kept their dependences on the final-state gluon momentum. 

It is now straightforward to contract this amplitude with the tree-level amplitude of the
single-emission process given in Eq.~\eqref{eq:M0R}.
Singling out the non-factorisable contributions, we obtain
\be
\begin{split} 
 2 {\rm Re} \big [  {\cal M}_0^*(1_q,2_b&,3_{q'},4_t;5_g) {\cal M}_{1}(1_q,2_b,3_{q'},4_t;5_g)  \big ]_{\rm nf} 
\\
&   = g_{s,b}^2 \; \frac{N^2-1}{4}
\left ( \frac{ \alpha_s}{2\pi} \right )
\left ( A_0^{L*}(5_g) \, B_1^{sH}(5_g) 
+ A_0^{H*}(5_g) \, B_1^{sL}(5_g) + {\rm c.c.} \right ).
\end{split} 
\label{eq:interf1R}
 \ee
 It follows from the definition of the colour-stripped amplitudes in Eq.~\eqref{eq:M1R}  that non-factorisable contributions
 are fully determined by Abelian amplitudes.

We are now able to discuss divergences and singular limits of non-factorisable amplitudes. Infrared divergences
 of  symmetric amplitudes $B_1^{sL(H)}$ do not depend on the fact that an additional gluon is emitted and, therefore, can still be
 described by the factor $I_1$ shown in Eq.~\eqref{eq:I1}. We find 
 \be
  B_1^{sL(H)}(1_q,2_b,3_{q'},4_t; 5_g) = \, I_1(\ep) \;  A_0^{L(H)}(1_q,2_b,3_{q'},4_t; 5_g) + B^{sL(H)}_{1, \rm fin }(1_q,2_b,3_{q'},4_t; 5_g) \, .
  \label{eq:RV_poles}
 \ee
 In addition to the infrared-divergent contribution to the one-loop, single-emission amplitude, we require its soft limit.
 Again, thanks to the Abelian nature of the amplitudes that contribute to non-factorisable corrections, we can write
 \be
 \begin{split}
   S_5 \, B_1^{sL}(1_q,2_b,3_{q'},4_t;5_g) &=  J(3,1;5,\varepsilon_5) \; B_1(1_q,2_b,3_{q'},4_t)\,,\\
   S_5 \, B_1^{sH}(1_q,2_b,3_{q'},4_t;5_g) &=  J(4,2;5,\varepsilon_5) \; B_1(1_q,2_b,3_{q'},4_t)\,.
\end{split} 
   \ee
   Hence,
\begin{align}
& S_5 \, \Big \{ 2 {\rm Re} \left [  {\cal M}_0^*(1_q,2_b,3_{q'},4_t;5_g) {\cal M}_{1}(1_q,2_b,3_{q'},4_t;5_g)  \right ]_{\rm nf} \Big \}
    \label{eq:RV_soft}
\\
&= -g_{s,b}^2 \;  \frac{N^2-1}{2} \left ( \frac{\alpha_s}{2 \pi} \right )\;
  {\rm Eik}_{\rm nf}(1_q,2_b,3_{q'},4_t; 5_g) \; 2 {\rm Re} \left [ A^*_0(1_q,2_b,3_{q'},4_t)  \; B_1(1_q,2_b,3_{q'},4_t)
    \right ]\nonumber \; , 
\end{align}
where the eikonal factor reads
\be
\label{eq:Eik_def}
{\rm Eik}_{\rm nf}(1_q,2_b,3_{q'},4_t ; k_g)  = J^\mu(3,1;k) J_\mu (4,2;k)
= 
\sum_{\substack{i \in [1,3]  \\  j \in [2,4] }}
     \frac{ \lambda_{ij} \; p_i \cdot p_j}{(p_i \cdot p_k) (p_j \cdot p_k)} \;,
   \ee
 with $\lambda_{ij} = +1 $ if both $i$ and $j$ are either incoming or outgoing, and $\lambda_{ij} = -1$ otherwise. 

 \subsection{Double-real emission amplitudes}

The double-emission process describes the radiation of two real gluons.
We parametrise this  process as follows
\be
\label{eq:RR_process_def}
1_q + 2_b \to 3_{q'} + 4_t + 5_g+6_g \;, 
\ee
and write the amplitude as
\be
\begin{split} 
&  \langle c |{\cal M}_0(1_q,2_b,3_{q'},4_t;5_g,6_g) \rangle  = 
 \\
 & \qquad
g_{s,b}^2 \Bigg [   \frac{1}{2} \{ t^{c_5}, t^{c_6} \}_{c_3 c_1} \delta_{c_4 c_2}  \, 
A_0^{sL}(5_g,6_g)
 +\frac{1}{2} [ t^{c_5}, t^{c_6} ]_{c_3 c_1} \delta_{c_4 c_2}  \, 
 A_0^{aL}(5_g,6_g)
\\
& \qquad \quad 
+\frac{1}{2} \{ t^{c_5}, t^{c_6} \}_{c_4 c_2} \delta_{c_3 c_1}  \, 
A_0^{sH}(5_g,6_g)
+\frac{1}{2} [ t^{c_5}, t^{c_6} ]_{c_4 c_2} \delta_{c_3 c_1}  \, 
A_0^{aH}(5_g,6_g)
\\
& \qquad \quad 
+t^{c_5}_{c_3 c_1}  t^{c_6}_{c_4 c_2} \,  B_0^{5L,6H}(5_g,6_g)
+t^{c_6}_{c_3 c_1}  t^{c_5}_{c_4 c_2} \,  B_0^{6L,5H}(5_g,6_g)
 \Bigg ].
\end{split} 
\label{eq:M0RR}
\ee
Similarly to Eq.~\eqref{eq:M1R}, we  split the full amplitude into amplitudes for emissions by
light- and heavy-quark lines. However, there are additional contributions when one gluon is emitted off  the light-quark line and the other off  the heavy-quark line.
Again, the colour-symmetric
parts are purely Abelian and the colour-antisymmetric ones are present because of  the non-Abelian nature of QCD.
Finally, we note that we have suppressed the dependence of the colour-stripped amplitudes on the quark momenta. 

It is straightforward to compute the non-factorisable contributions to the square of the
double-real emission amplitude shown in 
Eq.~\eqref{eq:M0RR}.
We account for  contributions such that each gluon is emitted and absorbed by a different quark
line and find 
\be
\begin{split} 
 \big|{\cal M}_0(1_q,2_b&,3_{q'},4_t;5_g,6_g)\big|_{\rm nf}^2 =g_{s,b}^4 \;  \frac{ N^2 - 1}{4} 
\\
& \times  \left( A_0^{sL}(5_g,6_g) \, A_0^{sH*}(5_g,6_g)
+  B_0^{5L,6H}(5_g,6_g) \, B_0^{6L,5H*}(5_g,6_g) + {\rm c.c.}  \right),
\label{eq5b}
\end{split} 
\ee
where the sum over colours has been performed and the sum over
polarisations of all final-state quarks and gluons is assumed.

It follows from Eq.~\eqref{eq5b} that since  the non-factorisable contributions  depend on particular combinations of 
colour-stripped amplitudes, they have peculiar properties. 
First, these contributions only depend on  the Abelian parts of the amplitudes.
Second, since only interference terms appear in  Eq.~\eqref{eq5b}, there are
no collinear singularities in the non-factorisable contributions.

We will need the single-soft limit of the double-real emission amplitude. Considering $p_6 \to 0$ as an example,
we  obtain the following soft limits of the colour-stripped amplitudes
\be
\begin{split} 
  S_6 \, B_0^{5L,6H}(1_q,2_b,3_{q'},4_t;5_g,6_g) & =   J(4,2;6,\varepsilon_6) \;  A_{0}^L(1_q,2_b,3_{q'},4_t;5_g)\;, \\
  S_6 \, B_0^{6L,5H}(1_q,2_b,3_{q'},4_t;5_g,6_g) & =    J(3,1;6,\varepsilon_6) \;  A_0^H(1_q,2_b,3_{q'},4_t;5_g)\;, \\
  S_6 \, A_0^{sL}(1_q,2_b,3_{q'},4_t;5_g,6_g) & =    J(3,1;6,\varepsilon_6)  \; A_0^{L}(1_q,2_b,3_{q'},4_t;5_g)\;, \\
    S_6 \, A_0^{sH}(1_q,2_b,3_{q'},4_t; 5_g,6_g) & =  J(4,2;6,\varepsilon_6)  \; A_{0}^{H}(1_q,2_b,3_{q'},4_t;5_g)\;.
\end{split} 
\ee
Hence,
\be
\begin{split}
  S_6 \,  \big|{\cal M}_0(1_q,2_b,3_{q'},4_t;5_g,6_g&)\big|_{\rm nf}^2  = -   g_{s,b}^4  \, \frac{N^2-1}{2}\;
  {\rm Eik}_{\rm nf}(1_q,2_b,3_{q'},4_t ; 6_g)
\\
 & \times \left [ A_0^{L}(1_q,2_b,3_{q'},4_t; 5_g) \, 
 A_0^{H*}(1_q,2_b,3_{q'},4_t; 5_g) + {\rm c.c.} \right ].
\end{split}
\label{eq:S6_RR}
\ee
We also need the double-soft limits of the colour-stripped amplitudes.
We make use of the fact that in an Abelian
theory soft limits of amplitudes fully factorise. Therefore, we obtain 
\be
\begin{split} 
  S_5 S_6  \, B_0^{5L,6H}(5_g,6_g) &  =  J(3,1;5,\varepsilon_5) \;  J(4,2;6,\varepsilon_6)  \, 
  A_0(1_q,2_b,3_{q'},4_t)\;, \\
   S_5 S_6 \, B_0^{6L,5H}(5_g,6_g) &  =   J(3,1;6, \varepsilon_6) \;  J(4,2;5,\varepsilon_5)  \, 
   A_0(1_q,2_b,3_{q'},4_t)\;, \\
   S_5 S_6  \, A_0^{sL}(5_g,6_g)   &  =   J(3,1;6,\varepsilon_6) \; J(3,1;5,\varepsilon_5)  \, 
   A_0(1_q,2_b,3_{q'},4_t)\;, \\
   S_5 S_6  \, A_0^{sH}(5_g,6_g)  &  =  J(4,2;6,\varepsilon_6) \;  J(4,2;5,\varepsilon_5)  \, 
   A_0(1_q,2_b,3_{q'},4_t)\;.
\end{split}
\ee
The double-soft limit of the non-factorisable contribution to the amplitude follows immediately.
We find 
\be
\begin{split} 
    S_5 S_6 \, \big|{\cal M}_0(1_q,2_b,3_{q'},4_t;5_g,6_g)\big|_{\rm nf}^2
  &= g_{s,b}^4\, (N^2 - 1)\,|A_0(1_q,2_b,3_{q'},4_t)|^2
  \\
&\;\; \times   {\rm Eik}_{\rm nf}(1_q,2_b,3_{q'},4_t; 5_g)  \; {\rm Eik}_{\rm nf}(1_q,2_b,3_{q'},4_t; 6_g)\,. 
\end{split} 
\label{eq:S5S6_RR}
\ee
   
\section{Construction of the subtraction terms}
\label{sec:pole_canc}

We can use the results of the previous section to extract singularities from non-factorisable contributions to single-top production.
For the sake of definiteness, we focus on the total cross section,
but the described procedure  applies verbatim to  any  {\it infrared-safe} observable.

\subsection{Double-real cross section}
We start by considering the process in Eq.~\eqref{eq:RR_process_def}, which we will refer to as the double-real contribution. To describe how the corresponding cross section
can be computed, we make use of the notation introduced in Ref.~\cite{Caola:2017dug} and define the function
\be
\label{eq:FLM_RR}
\begin{split}
	\flm^{\rm nf}\left(1_q,2_b,3_{q'},4_t;5_g,6_g\right) = \,  &
	{\cal N} 
	\int {\rm dLips}_{34} \, 
	(2\pi)^d \, \delta^{(d)} \Big(p_1+p_2 - \sum_{i=3}^{6} p_i \Big) 
	\\ & \qquad 
	\times \; \big|{\cal M}_0 \left(1_q,2_b,3_{q'},4_t;5_g,6_g\right) \big|_{\rm nf}^2 \; .
\end{split}
\ee
Here ${\rm dLips}_{34}$ is the Lorentz-invariant phase space of the two final
state fermions and $\cal N$ includes spin and colour averaging factors, ${\cal N} = 1/(4N^2)$. 
The total cross section can be
obtained by integrating over the phase space of the two gluons, $5_g$ and $6_g$,
and including
the appropriate symmetry factor. We write
\begin{align}
    \begin{split}
\label{eq:sigma_RR_def}
  2 s \cdot \sigma^{\rm nf}_{\text{RR}} & = 
 \frac{1}{2 !}  \int \measure{5} \measure{6} \flm^{\rm nf}\left(1_q,2_b,3_{q'},4_t;5_g,6_g\right)\\
                                         & \equiv \big\langle \flm^{\rm nf}\left(1_q,2_b,3_{q'},4_t; 5_g,6_g\right) \big\rangle \; , 
   \end{split}
\end{align}
where $s=2p_1\cdot p_2$ is the partonic centre-of-mass energy squared. The phase space element $\measure{}$ is defined as in Ref.~\cite{Caola:2017dug} and reads
\be
\measure{} = \frac{{\rm d}^{d-1}p}{(2\pi)^{d-1} 2E_p} \, \theta \big( E_{\rm max}-E_p \big) 
\label{eq:PS_element} \, ,
\ee
where $E_{\rm max}$ is a parameter that should be equal to or greater than the maximal
energy that a final-state parton can reach according to momentum conservation. In the present paper we use $E_{\rm max} = \sqrt{s}/2$.
The matrix element appearing in Eq.~\eqref{eq:FLM_RR} develops
singularities when at least one gluon becomes soft.
As we have already mentioned, no collinear divergences affect non-factorisable corrections since they are, essentially, the 
interference contributions. In order to preserve the fully differential nature of the calculation, 
we need to regulate and extract soft singularities without integrating over the resolved part of phase space.
To do so, we introduce the identity
\begin{align} 
 \la  \flm^{\rm nf}\left(1_q,2_b,3_{q'},4_t;5_g,6_g\right) \ra
  = \, &
 \la S_5 S_6 \, \flm^{\rm nf}\left(1_q,2_b,3_{q'},4_t;5_g,6_g\right) \ra 
\nonumber 
\\
& + 2\,  \la  S_6 \,  ( I- S_5) \, \flm^{\rm nf}\left(1_q,2_b,3_{q'},4_t;5_g,6_g\right) \ra
\\
& +\la  (I- S_5) \,  (I-S_6) \, \flm^{\rm nf}\left(1_q,2_b,3_{q'},4_t;5_g,6_g\right) \ra \; .
\nonumber 
\end{align}
The first term corresponds to the double-soft limit, the second one to the single-soft limit.
The last term gives the hard contribution where all singularities are regulated.
Note that these terms are symmetric under the exchange of the two gluons due to the factorisation in
Eq.~\eqref{eq:S5S6_RR} and the integration in Eq.~\eqref{eq:FLM_RR}.

The fully regulated term can be computed numerically in four dimensions without further ado.
However, we still need to treat the soft-divergent terms.
When soft operators act on $\flm^{\rm nf}$ they impact both the squared matrix 
element and the momentum conserving delta function in its 
definition.
The latter becomes independent of the soft momenta.
The single- and double-soft limits of the double-real, non-factorisable
matrix elements are given in Eqs.~\eqref{eq:S6_RR} and~\eqref{eq:S5S6_RR}, respectively.
In both cases, the matrix element factorises into the  universal structure
$\rm Eik_{\rm nf}$, defined in Eq.~\eqref{eq:Eik_def}, 
and a lower multiplicity matrix element, which does 
not depend on the soft radiation.
We can then integrate over the unresolved momenta without 
affecting the kinematics of the resolved partons.
We perform this integration using dimensional regularisation, i.e.
in $d = 4 - 2 \ep$ dimensions.
To present the result of the integration, we find it convenient to define the function
\be
\label{eq:K_def}
g_{s,b}^2  \int \measure{k} {\rm Eik}_{\rm nf}(1_q,2_b,3_{q'},4_t ; k_g) \equiv 
\frac{\alpha_s}{2\pi} \left ( \frac{2 E_{\rm max}}{\mu} \right )^{-2\ep}
K_{\rm nf}(1_q,2_b,3_{q'},4_t; \ep) \; .
\ee
The function  $K_{\rm nf}(\ep) \equiv K_{\rm nf}(1_q,2_b,3_{q'},4_t; \ep)$
can be found  in Appendix~\ref{app:int-ct} where it is computed 
up to $\mathcal{O}(\ep^0)$ terms. 
We now extract the soft divergences and write 
\be
\label{eq:sigma_RR}
\begin{split} 
2 s \cdot \sigma_{RR} = \, &
\left ( \frac{\alpha_s}{2\pi} \right )^2 \frac{N^2-1}{2 N^2}
    \left ( \frac{2E_{\rm max}}{\mu} \right )^{-4\ep}  
    \la  K_{\rm nf}^2(\ep)\;  F_{\rm LM}(1_q,2_b,3_{q'},4_t) \ra
\\
& - \left( \frac{\alpha_s}{2 \pi } \right)
\frac{N^2-1}{2}
\left ( \frac{ 2 E_{\rm max}}{\mu} \right )^{-2\ep}
 \la K_{\rm nf}(\ep) \, (I-S_5) \; {\widetilde F}^{\rm nf}_{\rm LM}(1_q,2_b,3_q,4_t;5_g) \ra
\\
   & +  \langle (I-S_5) (I-S_6)  \;  F^{\rm nf}_{\rm LM}(1_q,2_b,3_{q'},4_t;5_g,6_g) \rangle. 
\end{split} 
\ee
In the above equation we have introduced a function to describe the tree-level
process,
\be
\label{eq:FLM_born}
\begin{split}
	\flm\left(1_q,2_b,3_{q'},4_t\right) = \, & {\cal N} 
	\int {\rm dLips}_{34} \, 
	(2\pi)^d \, \delta^{(d)} \big(p_1\!+\!p_2\!-\!p_3\!-\!p_4\big) \,
	\big|{\cal M}_0 \left(1_q,2_b,3_{q'},4_t\right) \big|^2.
\end{split}
\ee
We stress that in Eq.~\eqref{eq:sigma_RR} the function $K_{\rm nf}(\ep)$ appears inside angular brackets
emphasising its dependence on the kinematics of hard particles.
We also introduced a non-factorisable, single-gluon emission contribution ${\widetilde F}_{\rm LM}^{\rm nf}$.
The tilde stresses that this contribution is defined  in terms of colour-stripped amplitudes
\be
\label{eq:FLM_tilde_R}
\begin{split}
        {\widetilde F}_{\rm LM}^{\rm nf}(1_q,2_b,3_q,4_t;5_g)
         = \, & {\cal N}
        \int {\rm dLips}_{34} \,
        (2\pi)^d \, \delta^{(d)} \Big(p_1+p_2-\sum_{i=3}^5 p_i\Big)
        \\ &
        \times \; g_{s,b}^2 \left (  A_0^{L*}(1_q,2_b,3_q,4_t;5_g) \, A_0^{H}(1_q,2_b,3_q,4_t;5_g) + {\rm c.c.} \right )\;.
\end{split}
\ee
This distinction is useful because such interference terms emerge from soft limits of higher-multiplicity amplitudes,
but otherwise do not contribute to non-factorisable corrections due to colour conservation.

\subsection{Real-virtual cross section}

A similar calculation can be performed for the real-virtual contribution to the single-top production cross section, 
which refers to the one-loop, non-factorisable corrections to the 
process in Eq.~\eqref{eq:1emission_proc}. By analogy with Eq.~\eqref{eq:FLM_RR}, 
we define 
\be
\label{eq:FLV}
\begin{split}
	\flv^{\rm nf} \left(1_q,2_b,3_{q'},4_t;5_g\right)
	= \, & {\cal N} 
	\int {\rm dLips}_{34} \, 
	(2\pi)^d \, \delta^{(d)} \Big(p_1+p_2 - \sum_{i=3}^{5} p_i \Big) 
	\\ &
	\times \; 
	2 {\rm Re} \Big [  {\cal M}^*_0(1_q,2_b,3_{q'},4_t;5_g) \,
	{\cal M}_{1}(1_q,2_b,3_{q'},4_t;5_g)  \Big ]_{\rm nf} \; .
\end{split} 
\ee
We obtain the real-virtual cross section by integrating 
$\flv^{\rm nf}$ over the phase space of the gluon $5_g$. 
Again, thanks to the fact that the non-factorisable
corrections are, effectively, 
Abelian and no collinear singularities are present, we can extract 
all singularities related to the emitted gluon by simply subtracting its soft limit. 
The real-virtual contribution reads
\begin{align}
    2 s \cdot \sigma_{\text{RV}} &= \int\measure{5}\flv^{\rm nf} \left(1_q,2_b,3_{q'},4_t;5_g\right) \nonumber \\
    &=  
    \la S_5 \flv^{\rm nf} (1_q,2_b,3_{q'},4_t; 5_g) \ra 
    + \la (I - S_5) \flv^{\rm nf} (1_q,2_b,3_{q'},4_t; 5_g) \ra \; .
\label{eq:sigma_RV_ref}
\end{align}
The first term is soft-divergent in the radiation phase space,
and the corresponding singularities become manifest 
once the integration over $\measure{5}$ is performed. 
The second term is soft-regulated. 
We notice that both contributions contain explicit poles in $\ep$,
stemming from ${\cal M}_1$ which appears in the definition of $\flv^{\rm nf}$.
We first analyse the soft-divergent term. 
Using the results in Eqs.~\eqref{eq:RV_soft} and~\eqref{eq:K_def}
we extract and integrate the soft factor, which multiplies 
a four-point, one-loop contribution. In order to make all  
divergences explicit, we exploit
Eq.~\eqref{eq:B1_4point} and obtain 
\be
\label{eq:S5_RV}
\begin{split} 
 \la S_5 \, \flv^{\rm nf} \, &(1_q,2_b,3_{q'},4_t; 5_g) \ra =
\\
& - \left (\frac{\alpha_s}{2\pi} \right )^2 \, \frac{N^2-1}{N^2} \,\left ( \frac{2E_{\rm max}}{ \mu} \right )^{-2\ep}
\la  K_{\rm nf}(\ep) \, {\rm Re}[I_1(\ep)] \, F_{\rm LM}(1_q,2_b,3_{q'},4_t) \ra
\\
& - \left (\frac{\alpha_s}{2\pi} \right )^2
\frac{N^2-1}{2}  \left ( \frac{2E_{\rm max}}{ \mu} \right )^{-2\ep}
\la K_{\rm nf}(\ep) \, {\widetilde F}^{\rm nf}_{\rm LV,\rm fin}(1_q,2_b,3_{q'},4_t) \ra \; ,
\end{split}
\ee
where the last term is proportional to the finite remainder 
of the single-virtual correction to the elastic process, Eq.~\eqref{eq:process}.
In particular, we have introduced 
\be
\label{eq:F_LV_fin}
\begin{split}
{\widetilde F}^{\rm nf}_{\rm LV,fin}(1_q,2_b,3_{q'},4_t) = \, & \,
{\cal N}
\int {\rm dLips}_{34} \,
(2\pi)^d \, \delta^{(d)} \big(p_1+p_2-p_3-p_4\big)
\\ & \qquad \times \;
        2 {\rm Re} \Big[
A_0^*(1_q,2_b,3_{q'},4_t) B_{1,\rm fin}(1_q,2_b,3_{q'},4_t)
\Big] \; ,
\end{split}
\ee
which is free of singularities, 
both explicit and implicit.

We now turn to the soft-regulated term in the second line of Eq.~\eqref{eq:sigma_RV_ref}.
According to Eqs.~\eqref{eq:interf1R} and \eqref{eq:RV_poles},
it only contains explicit poles  
and can be cast into  the following form
\be
\label{eq:ImS5_RV}
\begin{split} 
  \la (I-S_5) \flv^{\rm nf} & (1_q,2_b,3_{q'},4_t;5_g) \ra
  =
\\
&\;\;\;  \left ( \frac{\alpha_s}{2\pi} \right ) \, 
\frac{N^2-1}{2} \, 
\la {\rm Re}[I_1(\ep)] \, 
 (I-S_5) \, 
 {\widetilde F}^{\rm nf}_{\rm LM}(1_q,2_b,3_{q'},4_t; 5_g)
\ra
\\ &
+  \left ( \frac{\alpha_s}{2\pi} \right ) \, 
\frac{N^2-1}{4} \, 
\la
 (I-S_5) \, 
  {\widetilde F}^{\rm nf}_{\rm LV, fin}(1_q,2_b,3_{q'},4_t; 5_g)
\ra \; ,
\end{split} 
\ee
with ${\widetilde F}^{\rm nf}_{\rm LM}(1_q,2_b,3_{q'},4_t; 5_g)$ given in Eq.~\eqref{eq:FLM_tilde_R}. The last contribution in the
above equation 
is related to the finite remainder of the one-loop, five-point amplitude 
through the definition  
\be
\begin{split}
{\widetilde F}^{\rm nf}_{\rm LV, fin}(1_q,2_b,3_{q'},4_t; 5_g) &= \, 
{\cal N}
\int {\rm dLips}_{34} \,
(2\pi)^d \, \delta^{(d)} \Big(p_1+p_2 - \sum_{i=3}^{5} p_i \Big)
\\ & \times 
g_{s,b}^2 \,  \Big(
A_0^{L*}(1_q,2_b,3_{q'},4_t; 5_g) B_{1,\rm fin}^{sH}(1_q,2_b,3_{q'},4_t; 5_g) 
\\ &\quad\!
 + A_0^{H*}(1_q,2_b,3_{q'},4_t; 5_g) B_{1,\rm fin}^{sL}(1_q,2_b,3_{q'},4_t; 5_g)
+{\rm c.c.}  \Big) \, .
\end{split}
\ee
We note  that the finite remainders $B_{1,\rm fin}^{sL}$ and $B_{1,\rm fin}^{sH}$ require a dedicated calculation
that is discussed in Sec.~\ref{subsec:loop_amp}.

\subsection{Double-virtual cross section}

In the previous sections we constructed subtraction terms for 
the double-real
and the real-virtual contributions to the cross section, $\sigma_{\rm RR}$ and $\sigma_{\rm RV}$ respectively. These subtraction terms were 
 integrated over the unresolved phase space 
resulting in $1/\ep$ poles.
Moreover, we isolated the divergent part of the real-virtual 
amplitude and made all the singularities affecting this contribution explicit.
These $\ep$ poles have to cancel against 
$\ep$ poles in  the double-virtual contributions, 
which follow from Eq.~\eqref{eq:double_virt_poles}.
The double-virtual cross section can be written as
\be
\label{eq:sigma_VV_ref}
\begin{split}
2 s \cdot \sigma_{\text{VV}} &= 
\la \flvv^{\rm nf} (1_q,2_b,3_{q'},4_t) \ra
\\ &= 
{\cal N} \int {\rm dLips}_{34} \,
(2\pi)^d \, \delta^{(d)} \Big(p_1+p_2 - p_3-p_4 \Big)
\\ &\,\,  \times
\bigg\{	
\big|{\cal M}_1(1_q,2_b,3_{q'},4_t) \big|_{\rm nf}^2
+ 2 {\rm Re} \Big[ {\cal M}^*_0(1_q,2_b,3_{q'},4_t)
	{\cal M}_2((1_q,2_b,3_{q'},4_t) \Big]_{\rm nf} \bigg\}
\\  &= 
    \left (\frac{\alpha_s}{2\pi} \right )^2 \, \frac{N^2-1}{4} \, 
     \bigg[
    \frac2{N^2}  \, 
    \Big\langle 
\big(  {\rm Re} \left [ I_1(\ep) \right ] \big)^2  \, \flm (1_q,2_b,3_{q'},4_t)  
  \Big\rangle
  \\  &\qquad\qquad\:
  + 2 \, \la 
  {\rm Re} \left [ I_1(\ep) \right ]  {\widetilde F}^{\rm nf}_{\rm LV,\rm fin}(1_q,2_b,3_{q'},4_t)
  \ra 
  + \la  \widetilde{F}^{\rm nf}_{\rm VV,\rm fin}(1_q,2_b,3_{q'},4_t)  \ra
  \bigg] \; ,
\end{split}
\ee
where the first term in the square brackets collects all $1/\ep^2$ poles, 
the second contribution only contains $1/\ep$ poles, while the last term
is finite. We note that ${\widetilde F}^{\rm nf}_{\rm LV,\rm fin}$ is defined in Eq.~\eqref{eq:F_LV_fin}
and that ${\widetilde F}^{\rm nf}_{\rm VV,\rm fin}$ reads
\be
\begin{split}
	& \, \widetilde{F}^{\rm nf}_{\rm VV,\rm fin} (1_q,2_b,3_{q'},4_t)=
{\cal N}
\int {\rm dLips}_{34} \,
(2\pi)^d \, \delta^{(d)} \big(p_1+p_2-p_3-p_4\big)\\
    & \qquad \times \, 
\bigg\{ \, \big|B_{1,\rm fin}(1_q,2_b,3_{q'},4_t)\big|^2 
+ 2 {\rm Re} \Big[A_0^*(1_q,2_b,3_{q'},4_t) B_{2,\rm fin} (1_q,2_b,3_{q'},4_t) \Big]  \bigg\} \; .
\end{split}
\label{eq:FVVfin}
\ee

\subsection{Pole cancellation}
To obtain a manifestly finite expression for the non-factorisable contribution
to the total cross section we need to sum the results in Eqs.~\eqref{eq:sigma_RR}, 
\eqref{eq:S5_RV}, \eqref{eq:ImS5_RV} and 
\eqref{eq:sigma_VV_ref}. It is convenient to write this sum as the combination
of three different terms corresponding to final states with different resolved  multiplicities. We write 
\be
\sigma_{\rm nf} = \sigma_{\rm nf}^{(2g)} + \sigma_{\rm nf}^{(1g)} +\sigma_{\rm nf}^{(0g)},
\label{eq:sigmanf}
\ee
where
\be
2 s \cdot  \sigma_{\rm nf}^{(2g)} =  \la (I- S_5) (I-S_6) \flm^{\rm nf}\left(1_q,2_b,3_{q'},4_t; 5_g,6_g\right) \ra 
\label{eq:sigma2g}
 \ee
 is the fully regulated double-real emission contribution.
 It can be directly implemented in a numerical program.
In order to present the single-real emission contribution, $\sigma_{\rm nf}^{(1g)}$,
 and the elastic contribution, $\sigma_{\rm nf}^{(0g)}$, we introduce the
 following function 
 \be
 {\cal W}(1_q,2_b,3_{q'},4_t) = \left ( \frac{2E_{\rm max} }{ \mu} \right )^{-2\ep} K_{\rm nf}(1_q,2_b,3_{q'},4_t; \ep) - 
 {\rm Re} [ I_1(1_q,2_b,3_{q'},4_t;\ep) ] \; .
 \ee
We point out that ${\cal W}$ does not contain any $\ep$ pole. 
In fact, the first term in the $\ep$-expansion of $K_{\rm nf}$
describes  a soft, wide-angle emission  and
assumes a  simple form
\be
\label{eq:soft}
K_{\rm nf}(1_q,2_b,3_{q'},4_t; \ep)
=
  \frac1\ep \; \log \left( \frac{p_1 \cdot p_4 \; p_2 \cdot p_3}{p_1 \cdot p_2 \; p_3 \cdot p_4 }\right)
        + \mathcal{O}(\ep^0) \; .
\ee
Such a pole is cancelled by the singularities
arising from single-virtual corrections. In particular, using  Eq.~\eqref{eq:I1},
we find 
\be
K_{\rm nf}(1_q,2_b,3_{q'},4_t; \ep)
- {\rm Re}\big[ I_1(1_q,2_b,3_{q'},4_t; \ep) \big]
= {\cal O}(\ep^0) \; .
\ee
The single-real emission contribution, which corresponds to the sum
of Eq.~\eqref{eq:ImS5_RV} and the second term in Eq.~\eqref{eq:sigma_RR}, is then equal to
\be
\begin{split} 
2 s \cdot \sigma_{\rm nf}^{(1g)} =  \, &
  -\left ( \frac{\alpha_s}{2 \pi} \right ) \frac{N^2-1}{2} \, 
  \la \mathcal{W}(1_q,2_b,3_{q'},4_t) (I-S_5) {\widetilde F}^{\rm nf}_{\rm LM}(1_q,2_b,3_{q'},4_t; 5_g) \ra
\\
 &   + \left ( \frac{\alpha_s}{2 \pi}  \right ) \frac{N^2-1}{4} \,  
 \la (I-S_5) {\widetilde F}^{\rm nf}_{\rm LV,\rm fin}(1_q,2_b,3_{q'},4_t; 5_g) \ra \; .
\end{split} 
\label{eq:sigma1g}
\ee
It is free of both explicit and implicit singularities.
Finally, the finite, elastic contribution becomes 
\be
\begin{split} 
	2 s \cdot \sigma_{\rm nf}^{(0g)} = \, & 
\left ( \frac{\alpha_s}{2 \pi} \right )^2 \frac{N^2-1}{2N^2}  \, 
\la \mathcal{W}^2(1_q,2_b,3_{q'},4_t) \, F_{\rm LM}(1_q,2_b,3_{q'},4_t) \ra 
\\
& - \left ( \frac{\alpha_s}{2 \pi} \right )^2 \frac{N^2-1}{2}  \, 
\la \mathcal{W}(1_q,2_b,3_{q'},4_t) \, {\widetilde F}^{\rm nf}_{\rm LV,\rm fin}(1_q,2_b,3_{q'},4_t) \ra 
\\
&
+ \left ( \frac{\alpha_s}{2 \pi} \right )^2 \frac{N^2-1}{4}  \, 
\la \widetilde{F}^{\rm nf}_{\rm VV,\rm fin} (1_q,2_b,3_{q'},4_t) \ra \; .
\end{split} 
\label{eq:sigma0g}
\ee
As a final remark in this section, it is worth noting that in the entire procedure described here,
the only amplitude which must be expanded to $\mathcal{O}(\ep)$ is $B_1(1_q,2_b,3_{q'},4_t)$
as it is needed to extract the two-loop finite remainder $B_{2,\rm fin}(1_q,2_b,3_{q'},4_t)$ in Eq.~\eqref{eq:B2}.

\section{Amplitude calculation}
\label{sec:amp_calc}

In this section we discuss the calculation of the amplitudes needed to compute the non-factorisable cross-section defined in Eq.~\eqref{eq:sigmanf}. The obtention of the three tree-level amplitudes is shortly described, followed by the one- and two-loop amplitudes.

\subsection{Tree-level amplitudes}

To compute real-emission amplitudes
we generate the relevant diagrams  with \texttt{QGRAF}~\cite{Nogueira:1991ex} and process them
in \texttt{FORM}~\cite{Kuipers:2012rf,Ruijl:2017dtg}.
As single-top production is facilitated by  the exchange of a $W$ boson,
all  massless quarks that appear in these amplitudes are left-handed.
This can be seen at the diagram level by using the anti-commutativity of $\gamma_5$ to move the spin projectors $P_L = \frac{1}{2} (1 - \gamma_5)$ from the $W$ vertices to the incoming massless fermions.
Using standard bracket notation from spinor helicity formalism,\footnote{For a review of spinor-helicity formalism, see e.g. Ref.~\cite{Mangano:1990by}. For the case of massive fermions, see also Ref.~\cite{Kleiss:1985yh}.} we write
\be
P_L\, u(p_i) = u_L(p_i) = \vert i ]\,,\quad \textrm{for light-like $p_i$.}
\ee
This fixes the
helicity of the three massless external fermions, while the outgoing massive top quark can be both left- and right-handed.
By decomposing the momentum of the top quark into two massless momenta
\be
p_4 = p_4^\flat + \frac{m_t^2}{2 n \cdot p_4}\, n\;,
\ee
the massive Dirac-conjugate spinor can be written as 
\be
\bar{u}_L(p_4) = \la 4^\flat \vert + \frac{m_t}{[n 4^\flat]} [ n \vert \quad \textrm{and} \quad \bar{u}_R(p_4) = [ 4^\flat \vert + \frac{m_t}{\la n 4^\flat \ra} \la n \vert\;.
\ee
With these definitions we can write tree-level helicity amplitudes as
\be
\begin{split}
A_0(1_q^L,2_b^L,3_{q'}^L,4_t^L) &= \frac{g_W^2}{t - m_W^2} \la 3 4^\flat \ra [2 1] \;, \\
A_0(1_q^L,2_b^L,3_{q'}^L,4_t^R) &=\frac{g_W^2}{t - m_W^2} \frac{m_t}{\la n 4^\flat \ra} \la 3 n \ra [2 1] \;,
\end{split}
\ee
where $t = (p_1 - p_3)^2$ and $g_W = 2 m_W/v$ is the weak coupling constant defined through the $W$ boson mass, $m_W$, and the Higgs field vacuum expectation value, $v$.
By choosing $n = p_3$, we can force the
latter amplitude to vanish; this, in turn
yields more compact results for higher-multiplicity amplitudes.
The tree-level amplitudes have been cross-checked against \verb|MadGraph5_aMC@NLO|~\cite{madgraph}.

\subsection{Loop amplitudes}
\label{subsec:loop_amp}
We computed the non-factorisable four-point, one-loop amplitude $B_1(1_q,2_b,3_{q'},4_t)$ defined in Eq.~\eqref{eq:M1}, in 
an earlier  paper~\cite{Bronnum-Hansen:2021pqc}.
This amplitude enters the present calculation in both the real-virtual and double-virtual cross sections, see Eqs.~\eqref{eq:sigma_RV_ref} and~\eqref{eq:sigma_VV_ref} respectively.
We note that $\mathcal{O}(\ep)$ terms are only required for the latter for which we use the results obtained numerically for a fixed grid of phase space points in Ref.~\cite{Bronnum-Hansen:2021pqc}.
For the real-virtual contribution, amplitudes through ${\cal O}(\ep^0)$  are needed. 
To evaluate them,
we rely on \texttt{QCDLoops}~\cite{Ellis:2007qk,Carrazza:2016gav} for efficient and precise computation of one-loop integrals.

\begin{figure}[t]
    \centering
    \begin{subfigure}[ht]{0.3\linewidth}
        \centering
        \includegraphics[height=3.2cm]{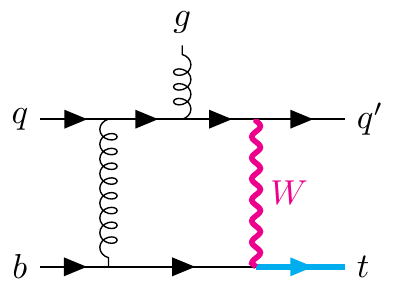}
    \end{subfigure}
    \hspace{2cm}
    \begin{subfigure}[ht]{0.3\linewidth}
        \centering
        \includegraphics[height=3.2cm]{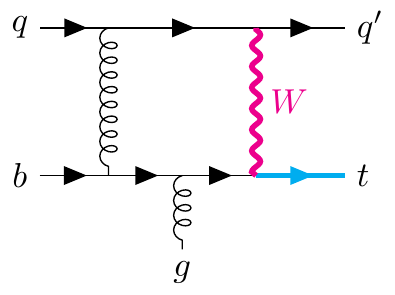}
    \end{subfigure}
    \caption{Examples of diagrams contributing to amplitudes $B_1^{sL}(5_g)$ (left) and $B_1^{sH}(5_g)$ (right) as defined through colour decomposition in Eq.~\eqref{eq:M1R}.}
    \label{fig:B1sX}
\end{figure}

The real-virtual contribution to the cross section also depends
on the one-loop, five-point amplitude $\mathcal{M}_1(1_q,2_b,3_{q'},4_t;5_g)$ defined in Eq.~\eqref{eq:M1R}.
The relevant colour-stripped amplitudes involve a gluon exchange between the fermion lines as well as a gluon emission.
The non-factorisable contribution to the cross section comes from interference between diagrams where the final-state gluon is
emitted and absorbed  by different  fermion lines.
Diagrams with a non-Abelian gluon vertex do not contribute to the cross section due to colour conservation, cf. Eq.~\eqref{eq:interf1R}.
For a straightforward extraction of the non-factorisable contribution, we calculate amplitudes for gluon emission from each of the two quark lines separately, $B_1^{sL}$ and $B_1^{sH}$.
In Figure~\ref{fig:B1sX} we present example diagrams for these two amplitudes.

A total of 24 diagrams contribute to $B_1^{sL}$ and $B_1^{sH}$.
We  generate them  with  \texttt{QGRAF} and process using  \texttt{FORM}.
We restrict external momenta to four dimensions, while loop momenta are considered to be $d$-dimensional.
Hence, the amplitude contains chains of Dirac matrices with $d$-dimensional indices between four-dimensional spinors.
The extra-dimensional part can be extracted by decomposing the matrices in four- and $(-2\ep)$-dimensional parts, $\gamma^\mu = \gamma^{\bar{\mu}} + \gamma^{\tilde{\mu}}$.
Indices with bars are restricted to four dimensions and indices with tilde are $(-2\ep)$-dimensional.
Spinor chains involving indices living in extra-dimensional space are projected on to tensors consisting solely of metric tensors restricted to the $(-2 \ep)$-dimensional subspace.
As an example, we write
\begin{align}
\ubar_t (p_4) \gamma^\mu \gamma^\nu u_b (p_2) = \ubar_t (p_4) \gamma^{\bar{\mu}} \gamma^{\bar{\nu}} u_b (p_2) + g^{\tilde{\mu} \tilde{\nu}}\, \ubar_t (p_4) u_b (p_2)\,.
\end{align}
After this procedure is applied,
all spinor chains involve objects with  four-dimensional indices so that  helicity amplitudes can be computed  straightforwardly.
As we already mentioned, due to the $W$ boson vertex all massless fermions are left-handed, hence there is a total of four helicity configurations.
The step of dimension splitting and helicity configuration projection is handled by the \texttt{FORM} library \texttt{spinney}~\cite{Cullen:2010jv}.

At this stage $B_1^{sL}$ and $B_1^{sH}$ can be written as linear
combinations of Feynman integrals, $I$, weighted by coefficients $c$. We write
\be
\begin{split}
B_1^{sX} = &\sum_{i} \sum_{r=0}^3 c^X_{5,i,r}(\ep)\,  I_{5,i} [k^{\mu_1} \cdots k^{\mu_r}] + \sum_{i} \sum_{r=0}^2 c^X_{4,i,r}(\ep) \, I_{4,i} [k^{\mu_1} \cdots k^{\mu_r}]\;,
\label{eq:B1sXstage1}
\end{split}
\ee
where $X = L,H$ and index $i$ labels the integral topology.
The coefficients acquire dependence on space-time dimension because of the dimension-splitting procedure described above.
For brevity, we have suppressed their dependence on kinematic invariants, four-dimensional spinor structures, and the electroweak coupling.
The integrals $I_{n,i}$ include pentagons ($n=5$) of up to rank 3 and boxes ($n=4$) of up to rank 2. We write 
\be
I_{n,i} [k^{\mu_1} \cdots k^{\mu_r}] = \int \frac{\mathrm{d}^{d}k}{(2\pi)^d} \, \frac{\prod_{j=1}^r k^{\mu_j}}{\prod_{l=1}^n D_{i,l}}\;,
\ee
where $D_{i,l} = (k - q_{i,l})^2 - m_{i,l}^2$ and the $q_{i,l}$ are given by sums of external momenta.
The propagator masses $m_{i,k}$ are zero, $m_t$, or $m_W$.

The most complicated integrals in Eq.~\eqref{eq:B1sXstage1} are tensor pentagon integrals of rank 3.
We reduce them to boxes of rank 2 and scalar pentagons by expanding the integrand numerator using the
van Neerven-Vermaseren (vNV) basis~\cite{vanNeerven:1983vr}.
We note that up to rank 3, pentagon integrals are free of rational terms~\cite{Bern:1994cg} and the expansion of the loop momentum in four dimensions is sufficient to obtain correct results in the $d \to 4$ limit.
Hence, we expand the loop momentum as
\begin{align}
k^\mu = \sum_{i=1}^4 (k \cdot q_i)\, v_i^\mu\,,
\end{align}
where we have used the fact that scalar products of vNV  basis vectors $v_i$ and propagator momenta $q_j$ satisfy
$v_i \cdot q_j = \delta_{ij}$. The
scalar products  $(k \cdot q_i)$ can be written in terms of kinematic invariants and inverse propagators.
We also note that $v_i^2 \neq 0$.
Applying this procedure, we obtain
\be
\begin{split}
B_1^{sX} = &\sum_{i} \tilde{c}^X_{5,i,0}(\ep)\,  I_{5,i} + \sum_{i} \sum_{r=0}^2 \tilde{c}^X_{4,i,r}(\ep)\,  I_{4,i} [k^{\mu_1} \cdots k^{\mu_r}] \\
&\,+ \sum_{i} \sum_{r=0}^1 \tilde{c}^X_{3,i,r}(\ep)\,  I_{3,i} [k^{\mu_1} \cdots k^{\mu_r}] + \sum_{i} \tilde{c}^X_{2,i,0}(\ep)\,  I_{2,i}\;,
\label{eq:B1sXstage2}
\end{split}
\ee
where we introduced the shorthand notation $I_{n,i}[1] \equiv I_{n,i}$.
At this stage we are left with scalar pentagon integrals and tensor integrals with at most four propagators.
Up to finite order in $\ep$, scalar pentagon integrals can be rewritten as boxes~\cite{Bern:1992em}.
The rest of the calculation employs the  Passarino-Veltman reduction~\cite{Passarino:1978jh} that allows
us to express the amplitude through  scalar integrals. We obtain 
\be
\begin{split}
B_1^{sX} = &\sum_{i} \hat{c}^X_{4,i}(\ep)\,  I_{4,i}+ \sum_{i} \hat{c}^X_{3,i}(\ep)\,  I_{3,i} + \sum_{i} \hat{c}^X_{2,i}(\ep)\,  I_{2,i} + \mathcal{O}(\ep)\;.
\label{eq:B1sXstage3}
\end{split}
\ee

After reduction the amplitude can be written in terms of 109 scalar box, triangle, and bubble integrals.
By switching to a basis with finite box integrals, the complexity of the integral coefficients reduces drastically.
We construct this basis following the ideas presented in Ref.~\cite{Badger:2016ozq}.
As an example of this basis change, we consider one of the box integrals
\begin{align}
I_{4,1} = \int \frac{\mathrm{d}^dk}{(2\pi)^d}\, \frac{1}{k^2 (k - p_1)^2 (k - p_1- p_2)^2 (k - p_1- p_2 + p_5 )^2}\;,\label{eq:I41}
\end{align}
that is infrared-divergent and the leading divergence is the second-degree $\ep$ pole. 
These singularities develop when one of the propagators goes on shell, for example when $k\to 0$ or $k \to p_1$.

We can regulate these singularities  by introducing an appropriate numerator in the integrand.
A suitable  numerator insertion  vanishes in the limits where the propagators that develop singularities go on shell.
For the integral in Eq.~\eqref{eq:I41} we use the following insertion
\begin{align}\label{eq:traceinsertion}
    \begin{split}
    \mathrm{tr}& \left( (-\slashed{p}_1) (\slashed{k} - \slashed{p}_1) (\slashed{k} - \slashed{p}_1 - \slashed{p}_2) (\slashed{p}_5) \right) = -s_{12}\, (s_{12} + s_{15} - s_{34}) + (s_{12} + s_{15} - s_{34})\, k^2 \\
&- (s_{12} - s_{34})\, (k - p_1)^2+(s_{12} + s_{15})\, (k - p_1 - p_2)^2 - s_{12}\, (k - p_1 - p_2 + p_5)^2\;.
\end{split}
\end{align}
We introduced $\slashed{p} = \gamma^\mu p_\mu$, as well as the usual Mandelstam variables
$s_{ij} = (p_i + \lambda_{ij} p_j)^2$ with $\lambda_{ij} = 1$ if the partons $i$ and $j$ are both incoming or outgoing
and $\lambda_{ij} = -1$ otherwise.
With this, we define the finite box integral
\begin{align}
F_{4,1} = \int \frac{\mathrm{d}^dk}{(2\pi)^d}\, \frac{\mathrm{tr} \left( (-\slashed{p}_1) (\slashed{k} - \slashed{p}_1) (\slashed{k} - \slashed{p}_1 - \slashed{p}_2) (\slashed{p}_5) \right)}{k^2 (k - p_1)^2 (k - p_1- p_2)^2 (k - p_1- p_2 + p_5 )^2} = \mathcal{O}(\ep^{0})\,.
\end{align}
It is clear from Eq.~\eqref{eq:traceinsertion} that the finite box is a linear combination of the divergent box and four triangle integrals.
Replacing all divergent boxes in the integral basis with
their finite counter-parts therefore
changes the triangle coefficients while leaving the box coefficients unchanged (up to an overall kinematic factor).
Hence, we obtain 
\be
\begin{split}
B_1^{sX} = &\sum_{i} \bar{c}^X_{4,i}\,  F_{4,i}+ \sum_{i} \bar{c}^X_{3,i}\,  I_{3,i} + \sum_{i} \hat{c}^X_{2,i}(\ep)\,  I_{2,i} + \mathcal{O}(\ep)\;.
\label{eq:B1sXstage4}
\end{split}
\ee

We note that the most complicated coefficients in the amplitude appear  in front of the finite box integrals.
However, since these integrals are finite we can  set $\ep \to 0$ in their coefficients. 
Furthermore, since in this basis $\ep^{-2}$ poles only appear in triangle integrals,  their coefficients must
be simple. 
In fact, after the procedure described above is applied,
the triangle integral coefficients either become independent of space-time dimension or simply vanish.
We note that due to the fact that the amplitude that we compute is UV-finite,
the sum of bubble integrals is  finite as well. 

One of the challenges in computing the real-virtual contribution to the cross section is that we need to compute
the amplitude in the limit when the emitted gluon is soft. 
To improve numerical stability in the evaluation of the amplitudes, we write the integral coefficients in terms of the kinematic invariants
\begin{align}
s_{12},\quad s_{23},\quad \delta_1 = s_{34} - s_{12},\quad \delta_2 = s_{45} - m_t^2,\quad \delta_3 = s_{15} \; .
\end{align}
In the limit where the emitted gluon goes soft, $p_5 \to 0$, the $\delta$-variables vanish.
Because of that, large cancellations in the integral coefficients can be avoided.

We have checked the real-virtual amplitude in
several ways. First, we compared its $\ep$ poles with  expectations shown in Eq.~\eqref{eq:RV_poles}.
We have also checked the factorisation in the limit where the emitted gluon goes soft, cf.  Eq.~\eqref{eq:RV_soft}.

As a final remark, we note that the
two-loop amplitude which is needed for the double-virtual contribution was computed in a previous
paper~\cite{Bronnum-Hansen:2021pqc}.
We used those results for the $q\,b \to q^\prime\,t$ channel
and obtained the contribution of the
$\overline{q}\,b \to \overline{q}^\prime\,t$ channel through the crossing symmetry, $p_1 \leftrightarrow -p_3$.
We note that no additional master integrals are required since after crossing the amplitude
can be mapped back to the basis of master integrals using integration-by-parts identities.

\section{Results}
\label{sec:results}
In this section we discuss the phenomenology of  the non-factorisable corrections to single-top production at the LHC. 
Our starting point is the conventional formula for the differential cross section 
\begin{align}
    {\rm d}\sigma_{pp \rightarrow X + t} = \sum_{i,j} \int\, {\rm d}x_1\,{\rm d} x_2\: f_i(x_1, \mu_F) \, 
    f_j(x_2, \mu_F)\:\,{\rm d} \hat{\sigma}_{ij\rightarrow X +t}\left(x_1,x_2\right)
    \,,
\label{eq5.1}
\end{align}
where we sum over partons that participate in the hard scattering. 

We take the CKM matrix to be an  identity matrix and work in the five-flavour scheme. The top quark in the final
state is produced in the collisions of a bottom quark from a proton and a virtual $W$ boson.
Overall,  the cross section in Eq.~(\ref{eq5.1}) receives contributions 
from  processes with  $i(j)=b$ and  $j(i)=u,c,\bar{d}, \bar{s}$.

We consider proton-proton collisions at 13~TeV and use the PDF set \texttt{CT14} in the computation.
We obtain the leading-order cross sections and distributions  using  the leading-order PDFs \texttt{CT14\_lo}
and the NNLO non-factorisable contribution  using the \texttt{CT14\_nnlo} PDF set.   
The strong coupling constant is provided by the  \texttt{CT14\_nnlo}  PDF set; numerically it evaluates to $\alpha_s(m_Z) = 0.118$. 
As the input parameters, 
we use  the vacuum expectation
value of the Higgs field,  $v= 246.2~{\rm GeV}$, the mass of the $W$ boson, $m_W = 80.379$ GeV,
and the pole mass of the top quark, $m_t = 173.0$ GeV.

The non-factorisable NNLO QCD correction  to the single-top production cross section is found to be
\be
\frac{\sigma_{pp\rightarrow X + t}}{1~{\rm pb}} = 117.96 + 0.26\left(\frac{\alpha_s(\mu_R)}{0.108}\right)^2
    \,,
\label{eq5.2}
    \ee
    where the first term on the right-hand side is the LO cross section
      and the second is the NNLO non-factorisable
      correction.\footnote{We stress one more time that the
       LO cross section in Eq.~\eqref{eq5.2} is computed with the LO PDFs and the 
        NNLO correction is computed with the NNLO PDFs.} To compute the cross sections shown in Eq.~\eqref{eq5.2}, we 
have set the factorisation scale to $\mu_F = m_t$. 
\begin{figure}[t]
    \centering
    \includegraphics[width=0.85\linewidth]{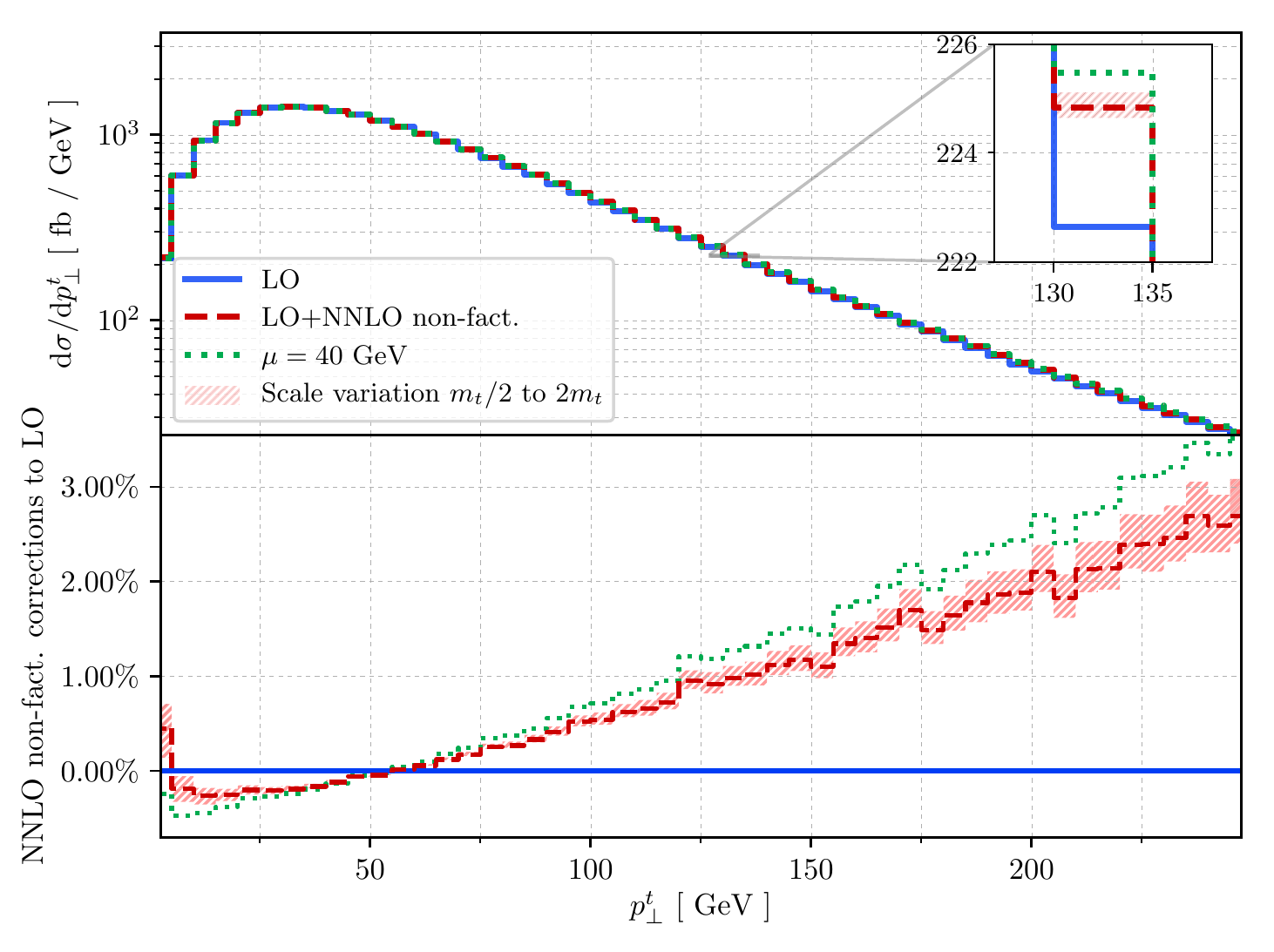}
    \caption{Distribution of the top-quark transverse momentum. The LO distribution is marked with a blue, solid line, while the red, dashed line corresponds to our predictions for
	the LO+NNLO distribution at $\mu=m_t$. The scale is variated between $\mu=m_t/2$ and $\mu=2m_t$. The green, dotted line corresponds to the scale $\mu=40$ GeV.
    The lower pane shows the ratio of non-factorisable corrections to the LO distribution. See text for further details.}
    \label{fig:pttop}
\end{figure}

We note however, that  since the 
non-factorisable contributions are absent at NLO due to colour conservation,
we do not have any indication of an optimal scale choice.  To emphasise this point, we have left
the dependence on the renormalisation scale explicit in
Eq.~(\ref{eq5.2}).
We note that the value of the strong coupling constant used there, $\alpha_s = 0.108$, corresponds to $\mu_R = m_t$
and for this choice of the scale 
the non-factorisable correction is about $0.2\%$ of the LO cross section .

However, it is unclear whether  $\mu_{R,F} = m_t$ is the right choice for the scales. Indeed, a typical
momentum transfer in $t$-channel single-top production is $\sim 40~{\rm GeV}$ since this is the value for which the
top-quark transverse momentum distribution is maximal. If we choose 
$\mu_R=40 \, {\rm GeV}$,  the non-factorisable corrections to the leading-order cross section become  $0.35\%$.
We note that this result  is in line with the recently published finite part of the  virtual contribution~\cite{Bronnum-Hansen:2021pqc}, which was found to be around  $0.5\%$ for  the same scale choice.

In what follows, we take $\mu_F = \mu_R \equiv \mu$, choose  $\mu = m_t$ as the central scale 
and estimate the scale uncertainty by increasing and decreasing  $\mu$ by a factor of 2.
As we already mentioned, even if we do not consider this choice to be optimal, it
does facilitate a comparison with
the literature, as  it was  used in several studies of
{\it factorisable} corrections to single-top production~\cite{Brucherseifer:2014ama, Campbell:2020fhf}.
In addition for a more realistic assessment of the magnitude of non-factorisable corrections, 
we also show  their impact for  $\mu=40\: \text{GeV}$.

We first present results for the top-quark transverse momentum 
distribution, see Figure~\ref{fig:pttop}.
It follows that non-factorisable corrections
are $p_{\bot}^t$-dependent; they are relatively small and negative at low values of the transverse momentum,
vanish at  $p_{\bot}^t  \sim {\cal O}(50~{\rm GeV})$ and reach 
${\cal O}(2\%)$ at $p_\bot^t\sim 200$~GeV.   This behaviour
is compatible with the fact that virtual contributions are negative in the
same $p_{\bot}^t$ interval~\cite{Bronnum-Hansen:2021pqc} and, as we explained in the introduction, virtual contributions to non-factorisable
corrections are expected to be dominant.
We note that shapes of   factorisable and non-factorisable corrections to the $p_{\bot}^t$ distribution are similar
but, typically, the factorisable ones are larger by a factor between $3$ and $10$, (see Figure 11 in Ref.~\cite{Campbell:2020fhf}). However, it follows from the same figure, that
the factorisable corrections vanish around $p_\bot^t \sim 30 \, {\rm GeV}$ 
whereas the non-factorisable ones vanish around $p_\bot^t \sim 50 \, {\rm GeV}$.
Hence, the non-factorisable corrections are, in fact, comparable to the factorisable ones in the region around
the \emph{maximum} of the $p_{\bot}^t$ distribution. 

\begin{table}
\centering
\begin{tabular}{@{} ll|ll|ll @{}}    \toprule
     & & \multicolumn{2}{c}{$\mu_R = m_t $}  & \multicolumn{2}{|c}{$\mu_R = 40\: {\rm GeV}$} \\
    $p_{\perp}^{t,\text{cut}}$ &  $\sigma_{\text{LO}}$ (pb) & $\sigma^{{\rm nf}}_{\text{NNLO}}$ (pb) & $\delta_{\text{NNLO}} \: [\%]$ & $\sigma^{{\rm nf}}_{\text{NNLO}}$ (pb) & $\delta_{\text{NNLO}}\: [\%]$\\ \midrule
 0 GeV & $118.01$  & $0.26_{+0.06}^{-0.04}$ & $0.22_{+0.05}^{-0.04} $ & $0.40$ & $0.34$ \\
 20 GeV & $115.09$  & $0.26_{+0.06}^{-0.04}$ & $0.23_{+0.05}^{-0.04} $ & $0.41$ & $0.36$ \\
 40 GeV & $109.56$  & $0.27_{+0.06}^{-0.05}$ & $0.25_{+0.06}^{-0.04} $ & $0.43$ & $0.39$  \\
 60 GeV & $104.63$  & $0.28_{+0.06}^{-0.05}$ & $0.26_{+0.06}^{-0.04} $ & $0.43$ & $0.41$ \\  \bottomrule

\hline
\end{tabular}
\caption{Dependence of the non-factorisable corrections on the top-quark 
transverse momentum. The factorisation scale is fixed to $\mu_F = m_t$. In the third column, the non-factorisable cross sections are evaluated at $\mu_R=m_t$ with sub- and super-scripts indicating the scale variation, $m_t/2$ and $2m_t$ respectively. The penultimate column describes the non-factorisable corrections at $\mu_R = 40\, {\rm GeV}$. For each scale choice, we report the relative impact, $\delta_{\text{NNLO}}$, of the non-factorisable contributions with respect to the LO cross section.}
\label{tab:pttop}
\end{table}

In Table~\ref{tab:pttop} we report the LO cross sections and 
the corresponding NNLO corrections  for different cuts on the minimal top-quark
transverse momentum. We fixed the factorisation scale to $\mu_F=m_t$ and inspect different renormalisation scales.
For $\mu_R=m_t$, we notice that, while the LO cross section {\it decreases}  by ${\cal O}(11\%)$ if the 
$p_\bot^t$ cut increases  from $0$ to $60$ GeV, the non-factorisable contribution to the 
cross section {\it increases}  by  ${\cal O}(8\%)$.
To understand the relative importance  of factorisable and non-factorisable NNLO corrections,
we note that factorisable corrections were computed to be about $-0.7\%$ of the NLO
cross section for similar choices of scales and parton distribution functions~(see Table 7 in Ref.~\cite{Campbell:2020fhf}).\footnote{Computations
  in Ref.~\cite{Campbell:2020fhf} were performed for proton-proton collisions at $14$~TeV.}
If we compare
this result with the fourth column of Table~\ref{tab:pttop} we  conclude that
the impact of non-factorisable corrections is smaller than, but quite comparable to, the factorisable corrections.
At $\mu_R=40 \, {\rm GeV}$, the NNLO non-factorisable corrections increase by $\mathcal{O}(8\%)$ by imposing a lower cut of $60 \, {\rm GeV}$ on the transverse momentum of the top quark.

\begin{figure}[t]
    \centering
      \begin{subfigure}{0.495\textwidth}
    \centering
    \includegraphics[width=\textwidth]{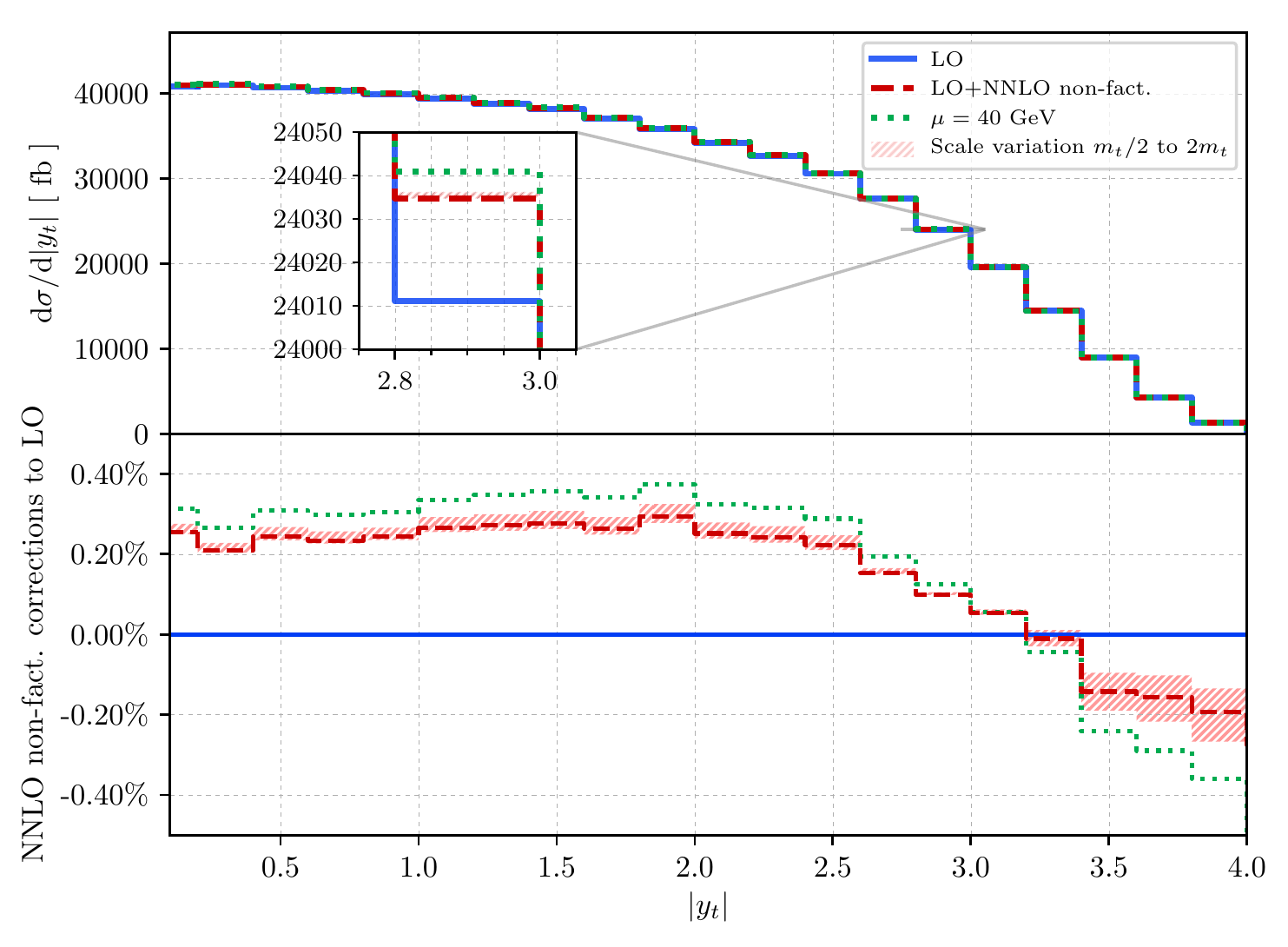}
    \caption{Distribution of the top-quark rapidity.}
    \label{fig:ytop}
  \end{subfigure}
  \begin{subfigure}{0.495\textwidth}
    \centering
    \includegraphics[width=\textwidth]{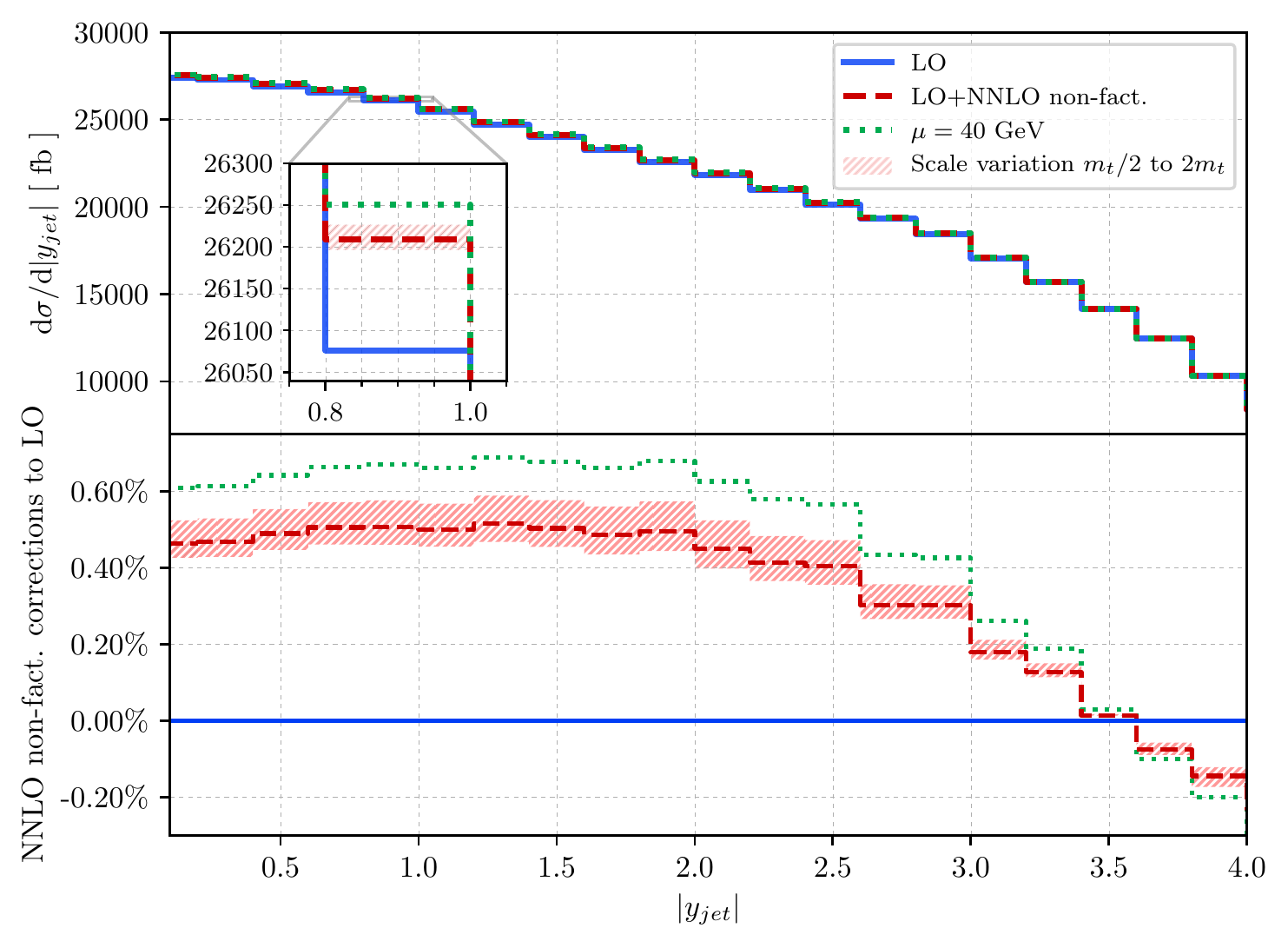}
    \caption{Distribution of the leading-jet rapidity.}
    \label{fig:yjet}
  \end{subfigure}
  \caption{Distributions of the absolute value of the top-quark rapidity (left) and of the leading-jet rapidity (right). LO distributions are marked with a 
	blue, solid line, while red, dashed lines correspond to our predictions for
	LO+NNLO distributions at $\mu=m_t$. The scale is variated between $\mu=m_t/2$ and $\mu=2m_t$. The green, dotted line corresponds to the scale $\mu=40$ GeV.
    Lower panes show  the ratio of non-factorisable corrections to LO distributions. See text for further details.}
  \label{fig:ytopandyjet}
\end{figure}

The top-quark rapidity distribution is shown  in the left pane of  Fig.~\ref{fig:ytopandyjet}.
The  (relative) non-factorisable corrections are fairly flat in the interval 
 $|y_t| < 2.5$ and change the leading-order rapidity distribution by   $\mathcal{O}(0.25\%)$.
For larger rapidity values, the corrections decrease rapidly and change
sign  at $|y_{t}| \sim 3$.
It follows from Ref.~\cite{Campbell:2020fhf} that 
the factorisable corrections to the top-quark rapidity distribution change the sign earlier, at around $|y_{t}|=1.2$.
Again, for such rapidity values, the non-factorisable and factorisable corrections are quite comparable. 

We turn to the analysis of the impact of non-factorisable corrections on jet observables in single-top production.
We use the $k_t$-algorithm~\cite{Ellis:1993tq} to define jets.
Jets are required to have  transverse momenta larger than  $30\: \text{GeV}$
and a radius  $R=0.4$.

In Fig.~\ref{fig:yjet} we show  the impact of non-factorisable corrections on the leading-jet rapidity distribution.
The correction is about $0.5\%$ at  small rapidities, $|y_{jet}| < 2$. Similar to the case of
the top-quark rapidity distribution, the correction to leading-jet rapidity
decreases and changes sign at around $|y_{jet}| \sim 3.5$.

\begin{figure}[t]
    \centering
      \begin{subfigure}{0.495\textwidth}
    \centering
    \includegraphics[width=\textwidth]{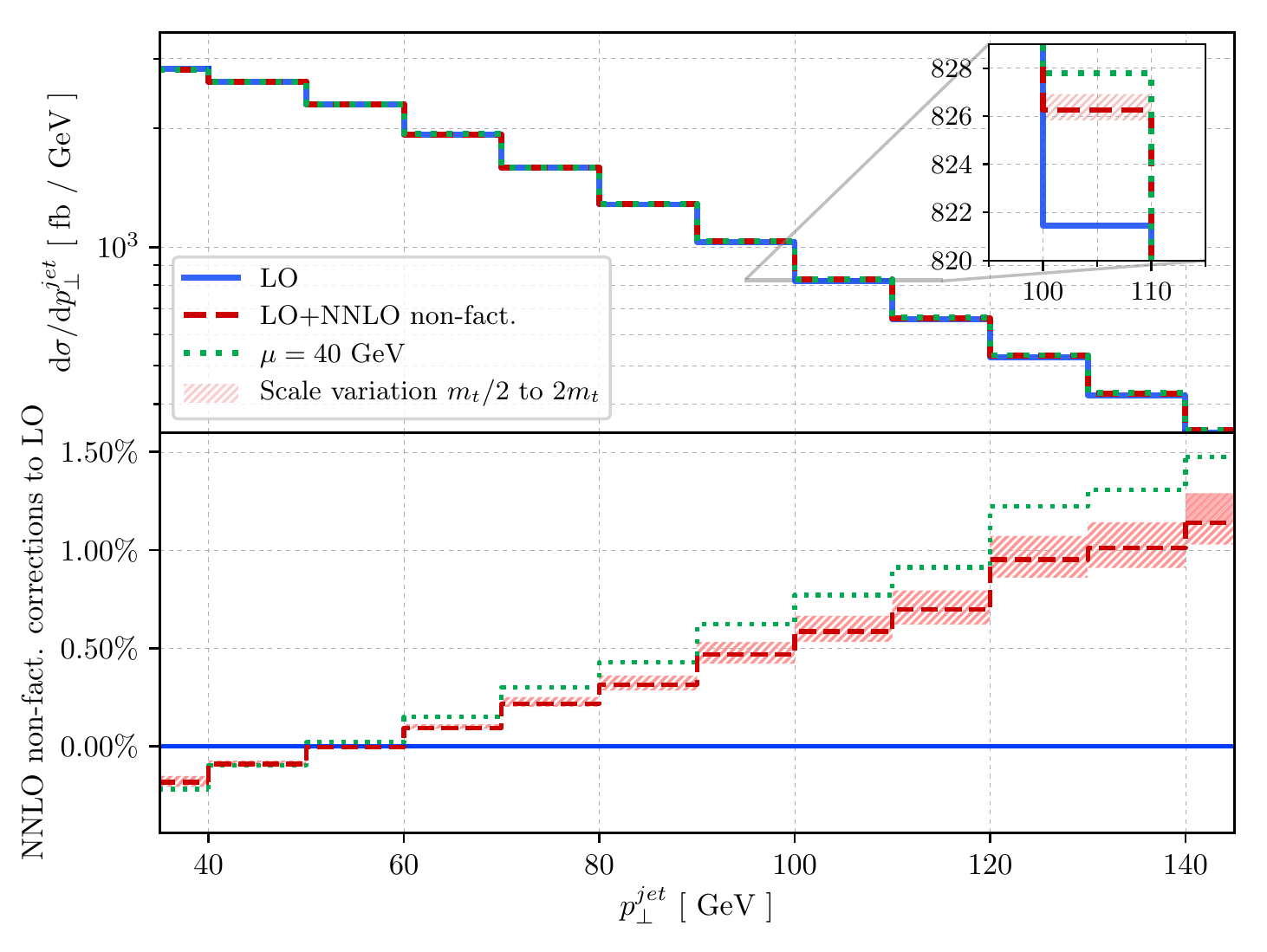}
    \caption{Distribution of the leading-jet transverse momentum.}
    \label{fig:ptjet}
  \end{subfigure}
  \begin{subfigure}{0.495\textwidth}
    \centering
    \includegraphics[width=\textwidth]{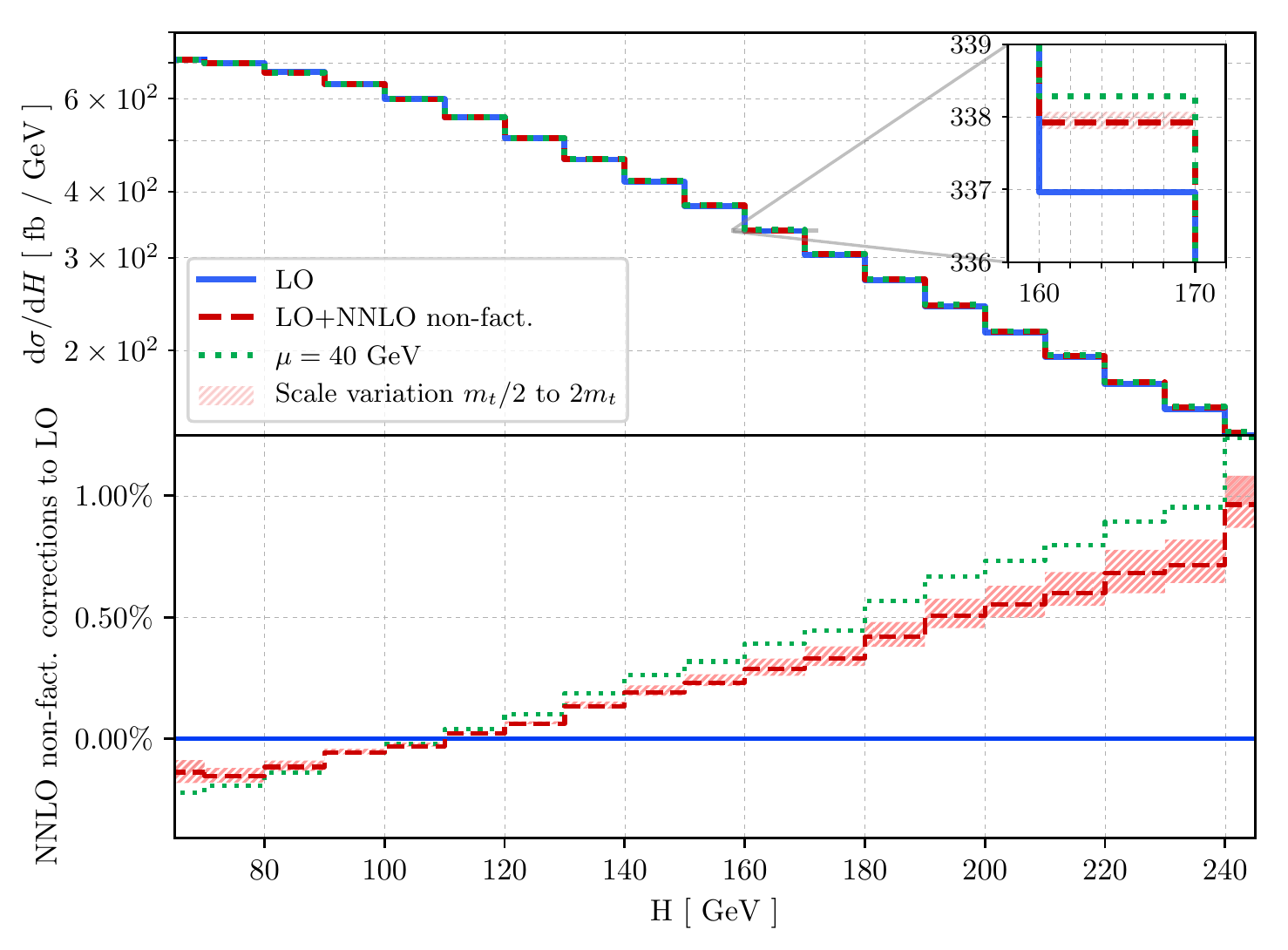}
    \caption{Distribution of the sum of transverse momenta $H$.}
    \label{fig:H}
  \end{subfigure}
  \caption{Distributions of the leading-jet transverse momentum (left)
      and of sum of transverse momenta  $H$ defined in Eq.~\eqref{eq:H} (right). LO distributions are marked with a
	blue, solid line, while red, dashed lines correspond to our predictions for
	LO+NNLO distributions at $\mu=m_t$. The scale is variated between $\mu=m_t/2$ and $\mu=2m_t$. The green, dotted line corresponds to the scale $\mu=40$ GeV.
    Lower panes show  the ratio of non-factorisable corrections to LO distributions. See text for further details.}
  \label{fig:ptjetandH}
\end{figure}

In Fig.~\ref{fig:ptjetandH} we show  the  transverse momentum distribution of the leading jet
(left pane) and  the distribution of the sum of the top and jets' transverse momenta,
\be
H_\perp = p^t_\bot + \sum_{i=1}^{n_{jet}} p^{\, jet, i}_\bot
    \, . 
    \label{eq:H}
\ee
In Eq.~(\ref{eq:H})  $n_{jet}$ is the number of reconstructed jets in an event.  The corrections to the leading-jet transverse momentum
distributions change sign  around $50 \, {\rm GeV}$, are negative for smaller $p^{\,jet}_\perp$ values and grow to about
$1.2$ percent at  $p_\bot^{ \, jet} \sim 140~{\rm GeV}$.
The distribution of the sum of transverse momenta
$H$ is affected by the non-factorisable corrections in a similar way.

\section{Conclusions}
\label{sec:conclusions}

In this paper we have computed the non-factorisable corrections  to
$t$-channel single-top production at the LHC. This contribution, being colour-suppressed
and computationally challenging, was neglected in all the previous studies of NNLO QCD
corrections to single-top production in spite of the recent indication that a peculiar enhancement
of such corrections  due to remnants of
the Glauber  phase is  possible~\cite{Liu:2019tuy}.

We have shown how to overcome the technical challenges related to the computation of virtual, non-factorisable
corrections in Ref.~\cite{Bronnum-Hansen:2021pqc}. In this paper we completed the calculation
of these corrections by including double-real and real-virtual contributions required to obtain the
infrared-finite cross section.  We
have discussed the calculation of the relevant  tree- and one-loop  amplitudes needed for the computation of non-factorisable corrections. 
Because  of the large number of mass scales that appear in the computation of one-loop amplitudes required for the real-virtual
non-factorisable contribution,
its reduction to master integrals, and its stable and efficient numerical evaluation turn out to be non-trivial. 
We discussed  how to address  these problems  and pointed out that it is beneficial to choose infrared-finite combinations
of boxes and triangles  as master integrals. 

We have explicitly shown that the non-factorisable corrections are not affected by the non-Abelian nature of QCD and are free of collinear singularities.
We have constructed subtraction terms  that make 
the cancellation of infra-red singularities in arbitrary infrared-safe observables explicit. 

We have studied  a number of  kinematic distributions relevant for  the single-top production process 
as well as  the inclusive cross section.
We have found that non-factorisable corrections are smaller than, but quite comparable to, the
factorisable ones.  Since the choice of the proper renormalisation scale in the non-factorisable corrections
is an open issue, the actual magnitude of these corrections is uncertain.
We estimate that they can reach ${\cal O}(0.4\%)$ in case of the inclusive cross section and ${\cal O}(1-2\%)$
for some kinematic distributions.   Another interesting point is that for many distributions the non-factorisable
corrections {\it do not} reduce to an overall renormalisation of the leading-order distributions. Thus, if a percent-level
precision in single-top studies can be reached, the non-factorisable effects will have to be taken into account. 

\acknowledgments
C. S-S. would like to thank Paolo Torrielli for useful discussions. 
This research is partially supported by the Deutsche Forschungsgemeinschaft (DFG, German Research Foundation) under grant 396021762 - TTR 257. The diagrams in Figure~\ref{fig:example} and \ref{fig:B1sX} were generated using \texttt{tikz-feynman}~\cite{Ellis:2016jkw}.

\appendix

\section{Renormalisation}
\label{sec:app_A}
Since the non-factorisable NLO QCD corrections vanish due to colour conservation, there is no UV divergences at NNLO. Nevertheless, the coupling is renormalised in $\overline{\text{MS}}$ scheme to zeroth order in perturbative QCD
\begin{align}
\frac{g_{s,b}^2}{4 \pi}
\equiv 
    \alpha_s^{\text{bare}} = \mu^{2\ep}S_{\ep} \, \alpha_s\left(\mu\right)    \,,
\end{align}
where $S_{\ep} = \exp\left(\ep\gamma_E\right)/(4\pi)^\ep$ and $\gamma_E\approx 0.57721$ is the Euler-Mascheroni constant.

\section{Integrated counterterms}\label{app:int-ct}
In this section we describe the calculation of the single-soft 
integrated counterterms. As explained  in the main body of the paper, 
it is important for treating the infrared 
singularities that originate  from the real radiation. 
We have previously defined the function $K_{\rm nf}(\ep)$ through the following integral
\be
g_{s,b}^2  \int \measure{k} {\rm Eik}_{\rm nf}(1_q,2_b,3_{q'},4_t; k_g) 
=
\frac{\alpha_s}{2\pi} \left ( \frac{2E_{\rm max}}{\mu} \right )^{-2\ep} K_{\rm nf}(1_q,2_b,3_{q'},4_t; \ep)
\; .\label{eq:K_def_app}
\ee
Again, we stress that the coupling $g_{s,b}$ appearing on the left-hand side of Eq.~\eqref{eq:K_def_app} is the bare coupling constant, while on the right-hand side $\alpha_s$ is the coupling renormalised at the scale $\mu$.
The latter is obtained by using the prescription in Appendix~\ref{sec:app_A}. 
Moreover, we have defined 
\begin{align}
    \begin{split}
        {\rm Eik}_{\rm nf}(1_q, 2_b, 3_{q'}, 4_t; k_g&)
    =
	\sum_{\substack{ i \in [1,3]  \\ j \in[2,4]}} 
	\frac{\lambda_{ij} \, p_i\cdot p_j}{(p_i \cdot p_k) \, (p_j \cdot p_k)}
	\\ = \, &
	\frac{p_1\cdot p_2}{p_1 \cdot p_k \, p_2 \cdot p_k}
	-
	\frac{p_1\cdot p_4}{p_1 \cdot p_k \, p_4 \cdot p_k}
	- 
	\frac{p_2\cdot p_3}{p_2 \cdot p_k \, p_3 \cdot p_k}
        +
	\frac{p_3\cdot p_4}{p_3 \cdot p_k \, p_4 \cdot p_k} \, .
    \end{split}
\end{align}
We note  that the above expression  involves two different 
structures: eikonal factors that depend on the four-momenta of
two massless partons,   and eikonal factors that depend
on the four-momenta of one massive and one massless 
parton.
The integration over the unresolved radiation is different
in the two cases.

Before proceeding with the details of the calculation, 
we quote the final result in order to highlight its simplicity. The result reads
\be
\label{eq:softB}
\begin{split}
K_{\rm nf}(1_q,2_b,3_{q'},4_t; \ep) = & \,  
\frac1\ep \, \log \left( \frac{p_1 \cdot p_4 \; p_2 \cdot p_3}{p_1 \cdot p_2 \; p_3 \cdot p_4 }\right)
- \frac12 \log^2 \left( \frac{\rho_{23}}{2}\right)
+ \frac12 \log^2 \left( \frac{\rho_{12}}{2}\right)
\\
& 
- \log \left( \frac{\rho_{14}}{1-\beta}\right)
\log\left( \frac{\rho_{14}}{1+\beta} \right)
+ \log \left( \frac{\rho_{34}}{1-\beta}\right)
\log\left( \frac{\rho_{34}}{1+\beta} \right)
\\
&
+ {\rm Li}_2 \left( 1- \frac{\rho_{12}}{2}\right)
- {\rm Li}_2 \left( 1- \frac{\rho_{23}}{2}\right)
\\
&
- {\rm Li}_2 \left( 1- \frac{\rho_{14}}{1-\beta}\right)
- {\rm Li}_2 \left( 1- \frac{\rho_{14}}{1+\beta}\right)
\\
&
+ {\rm Li}_2 \left( 1- \frac{\rho_{34}}{1-\beta}\right)
+ {\rm Li}_2 \left( 1- \frac{\rho_{34}}{1+\beta}\right)
+ \mathcal{O}(\ep)
\; ,
\end{split}
\ee
where $\beta = \sqrt{1 - m_t^2/E_4^2}$~ and $\rho_{ij} = p_i\cdot p_j/( E_i E_j)$.
To simplify the discussion, it is convenient to define a function $I_{\Omega}$ that contains all the information about
integration over angles of the emitted gluon. We write 

\be
\begin{split}
g_{s,b}^2
\int  \measure{k} \frac{p_i\cdot p_j}{\left(p_i\cdot p_k\right)\left(p_j\cdot p_k\right)} 
= \, &
 \frac{\alpha_s}{2\pi} \,
 \Big( \frac{2E_{\rm max}}{\mu}\Big)^{-2\ep} \, 
\frac{e^{\ep \, \gamma_E} \Gamma(1-\ep)}{\Gamma(1-2\ep)} \, 
\Big( -\frac{1}{2\ep}\Big)
  \\
&  \qquad \times 
\int {\rm d} \cos \theta \; \frac{{\rm d} \phi}{\pi}
\big( \sin \theta  \sin \phi \big)^{-2\ep} \, 
\frac{\hat{p}_i \cdot \hat{p}_4}{\hat{p}_i \cdot \hat{p}_k \, \hat{p}_4 \cdot \hat{p}_k}
 \\
= \, &
\frac{\alpha_s}{2\pi} \,
 \Big( \frac{2E_{\rm max}}{\mu}\Big)^{-2\ep} \, 
\frac{e^{\ep \, \gamma_E} \Gamma(1-\ep)}{\Gamma(1-2\ep)} \, 
\Big( -\frac{1}{2\ep}\Big)
 \; I_{\Omega},
\end{split}
\ee
where $\hat{p} = p/E_p$.
We present the results for the functions $I_{\Omega}$ for the three relevant cases below.

\subsection{One massive and one massless emitter - arbitrary angle}

We consider the case when  one emitter is massless, $p_i^2 = 0$,  and the other is massive, $p_{4}^2 = m_t^2$. 
The function $I_{\Omega}$ reads 
\begin{align}
      I_{\Omega} &= -\frac{1}{\ep} 
      + \mathcal{I}^{(0)}
      + \ep \,  \mathcal{I}^{(1)}
      + \mathcal{O}\left(\ep^2\right)
    \,,
\end{align}
where the different terms reads
\be
\begin{split}
\mathcal{I}^{(0)} = \, & 2 \log \bigg(  \frac{E_4 \, \rho_{i4}}{m_t } \bigg) \; , 
 \\
\mathcal{I}^{(1)} = \, &
-2 \bigg[
\frac14
\log^2 \Big(  \frac{1-\beta}{1+\beta} \Big)
+ \log \Big( \frac{\rho_{i4}}{1+\beta}  \Big)\; 
 \log \Big(  \frac{\rho_{i4}}{1-\beta} \Big) 
 \\ & \qquad 
+{\rm Li}_2 \Big( 1- \frac{\rho_{i4}}{1+\beta} \Big)
+{\rm Li}_2 \Big( 1- \frac{\rho_{i4}}{1-\beta} \Big)
 \bigg].
 \end{split}
\ee
 The explicit expressions for $\mathcal{I}^{(0)}$ and $\mathcal{I}^{(1)}$
 agree with the results in Ref.~\cite{Alioli:2010xd}.

   \subsection{One massive and one massless emitter - back-to-back kinematics}
   The previous result
   simplifies when the two emitters are back-to-back.  We consider the case of one massless and one massive emitter, 
   $p_i^2=0$ and $p_4^2=m_t^2$, in the case when $\vec{p}_i + \vec{p}_4 = 0$. The function $I_{\Omega}$ reads 
   \begin{align}
       I_{\Omega}
       = \, &
        2^{-4\ep} \, B\left(\frac{1}{2}-\ep,\frac{1}{2}-\ep\right)
        \\ 
        & \, \times
       \bigg[B\left(-\ep,1-\ep\right) + \frac{2\beta}{1+\beta}\frac{\Gamma^2\left(1-\ep\right)}{\Gamma\left(2-2\ep\right)}\:{}_2F_1\left(1,1-\ep,2-2\ep,\frac{2\beta}{1+\beta}\right)\bigg]
        \,.
           \end{align}
   \subsection{Two massless emitters - arbitrary angles}
When the two  emitters are massless, the result reads (see e.g. Ref.~\cite{Caola:2017dug})
\be
I_{\Omega}
= 
-
\frac{1}{\ep}   \, 
\rho_{ij} \; 
{}_2F_1 \Big(1, 1, 1-\ep, 1-\frac{\rho_{ij}}{2}\Big) \, .
\label{eq:soft_int_proof}
\ee

\bibliographystyle{JHEP}
\bibliography{references}

\end{document}